\documentclass{aastex63}

\usepackage{ulem}
\usepackage{xcolor}
\usepackage{amsmath}
\usepackage{amssymb}
\usepackage{lineno}
\usepackage{placeins}
\usepackage{natbib}
\submitjournal{\apj}
\accepted{05 Oct 2020}

\begin{document}

\title{The Rapid Variability of Wave Electric Fields within and near quasiperpendicular interplanetary shock ramps: STEREO Observations}

\author{Z.A. Cohen}
\affiliation{School of Physics and Astronomy, University of Minnesota, Minneapolis, Minnesota, USA.}

\author{C. A. Cattell}
\affiliation{School of Physics and Astronomy, University of Minnesota, Minneapolis, Minnesota, USA.}

\author{A. W. Breneman}
\affiliation{School of Physics and Astronomy, University of Minnesota, Minneapolis, Minnesota, USA.}

\author{L. Davis}
\affiliation{School of Physics and Astronomy, University of Minnesota, Minneapolis, Minnesota, USA.}

\author{P. Grul}
\affiliation{School of Physics and Astronomy, University of Minnesota, Minneapolis, Minnesota, USA.}

\author{K. Kersten}
\affiliation{School of Physics and Astronomy, University of Minnesota, Minneapolis, Minnesota, USA.}

\author{L. B. Wilson III}
\affiliation{NASA Goddard Space Flight Center, Greenbelt, Maryland, USA.}

\author{J. R. Wygant}
\affiliation{School of Physics and Astronomy, University of Minnesota, Minneapolis, Minnesota, USA.}

\correspondingauthor{Zachary Cohen}
\email{cohen512@umn.edu}

%\begin{keypoints}
%\item Highly variable wave modes and amplitudes from 5 to 200 mV/m are observed near ramps
%\item First observations of waves in ion acoustic frequency range with frequency drift are reported
%\item First observations of ECDI-like waves seen both upstream and downstream of shock ramps, in contrast to theoretical predictions
%\end{keypoints}

\begin{abstract}
We present STEREO observations within 1500 proton gyroradii of 12 quasiperpendicular interplanetary shocks, with long-duration burst mode electric field acquisition by S/WAVES enabling observation of the evolution of waves throughout the entire ramp of interplanetary shocks. The shocks are low Mach number ($M_{f} \sim $1--5), with beta ($\beta$) $\sim $0.2--1.8. High variability in frequency, amplitude, and wave mode is observed upstream, downstream, and in shock ramps. Observations in every region include ion acoustic-like waves, electron cyclotron drift instability driven waves, electrostatic solitary waves, and high frequency whistler mode waves. We also show for the first time the existence of electrostatic waves with frequencies in the ion acoustic range which are frequency dispersed in time and the first observations of electron cyclotron drift instability (ECDI) driven waves at interplanetary shocks. The waves are bursty, large amplitude ($\sim$5 to $>$ 200 mV/m), and seen in all three regions. All wave modes are more commonly observed downstream than upstream.
%, usually within $\sim$ 63000 km ($\sim 1500 \rho_{gi}$) of the ramp.
\end{abstract}

\section{Introduction}
Collisionless shock waves have the potential to heat and accelerate charged particles; waves are often proposed to provide heating. The solar wind contains many free energy sources, each of which can lead to the growth of specific electrostatic and/or electromagnetic wave modes. Often each wave mode is identified with the particular instability that converts free energy into the wave energy. Since shocks have rapid variations in plasma parameters (e.g., bulk flow speed, magnetic field magnitude and direction, particle density) there are many processes that contribute to changes in the particle distributions, including particle beams, particle reflections, and plasma waves. Observations of the particle distributions and the plasma waves being generated provide details of the microphysical mechanisms providing dissipation at shocks and of the physics of shocks in general.

The earliest in situ studies of collisionless shocks took place at Earth's bow shock, following predictions of the existence of bow shocks in the 1950s and 1960s, \citep[e.g.,][]{Gardner1958,S1958,Mor1961,K1962}.
The first low frequency ($\sim$0.56 -- 70 kHz) electric field measurements at Earth's bow shock were recorded by the OGO-5 spacecraft \citep{F1968}. Later, the ISEE \citep{O1977} spacecraft provided the first view of high-resolution electric field waveforms at the bow shock of the earth, as these were part of the pioneering mission to carry devices capable of measuring the time domain electric fields in short-duration, high-resolution bursts \citep{M1978}. Based upon the ongoing observations of these and other earlier research and experiments, researchers proposed that anomalous resistivity due to wave particle interactions could provide energy dissipation and heating necessary to initiate a shock from a nonlinearly steepening wave \citep[]{V1963,S1966,G1979,T1985}.  In particular, Langmuir waves \citep{T1929,B1949a,B1949b,F1968}, ion acoustic (or ion acoustic-like) waves \citep{F1968,F1970,Formisano1982}, lion roars \citep{Z1999,Giagkiozis2018} ,magnetosonic whistler mode waves \citep{F1974}, and other wave modes \citep[]{F1970,RG1975} were observed at the bow shock. Other mechanisms (e.g., ion reflection) which may not result in strong waves are also invoked \citep{P1980,Gary1981}.

Ulysses \citep{W1992}, WIND \citep{L1995}, and other investigations provided a further look at waves in the interplanetary medium \citep[e.g.,][]{L1998,H1998,H2006,W2007}. Interplanetary (IP) shocks, generated by interplanetary coronal mass ejections (ICMEs) or stream interaction regions (SIRs), are of particular interest because they provide a large database of collisionless, low (fast) Mach number ($M_{fast}\simeq$ 1--5, this study) shocks with $\beta$ ranging from $\simeq$ 0.2--1.8.  As in the case of the bow shock, anomalous resistivity due to wave-particle interactions is considered to be one of the primary energy dissipation mechanisms in IP shocks \citep[e.g.,][]{Gary1981,Fitz2003,W2014a,W2014b,M2016}, especially when the fast Mach number is below its critical value, $M_{cr}$, which is typically $\sim$ 1-2 in quasi-perpendicular ($\theta_{Bn} \geq 45^{\circ}$) interplanetary shocks \citep{C1970,K1987,EK1984,Kr2002}), where $\theta_{Bn}$ is the angle between the magnetic field and the shock normal direction.

%Many different wave modes have been predicted by simulation and/or observed near collisionless shocks, including Langmuir waves \citep[e.g.,][]{H1998,M1999,W2007}, 
%electrostatic solitary waves (ESWs or solitary waves) \citep{M1999,W2007}, 
%ion acoustic-like waves (IAWs or IA-like) \citep{K1982,H1998,M1999,W2007,Goodrich2018,Goodrich2019}, 
%electron cyclotron harmonics (Bernstein mode) possibly driven by the electron cyclotron drift instability (ECDI) \citep{ML2006,ML2013,ML2017,W2010,Bren2013,W2014a,W2014b,Goodrich2018},
%and electromagnetic whistler mode waves  \citep{K1982,Bren2010,W2009,W2012,W2013,W2017}.

Many different wave modes have been predicted by simulation and/or observed near collisionless shocks, including Langmuir waves, electrostatic solitary waves (ESWs or solitary waves), ion acoustic-like waves (IAWs or IA-like), electron cyclotron harmonics (electron Berstein mode) possibly driven by the electron cyclotron drift instability (ECDI), and electromagnetic whistler mode waves.\citep{K1982,H1998,M1999,ML2006,ML2013,ML2017,Bren2013,W2007,W2009,W2010,W2012,W2013,W2014a,W2014b,W2017,Goodrich2018,Goodrich2019}

The highest frequency, commonly observed waves, Langmuir waves, are electrostatic waves, linearly polarized parallel to the magnetic field, with narrow frequency peaks at or near the electron plasma frequency ($f_{pe}$), and are thought to be usually generated by the bump-on-tail instability \citep{B1949a,B1949b}. ESWs are nonlinear bipolar pulses in the electric field, mostly parallel to the magnetic field, and have been related to electron beams creating electron holes \citep{Bale1998}, or possibly ion holes \citep{Vasko2018}.

Three wave modes are observed with frequencies from the electron cyclotron frequency ($f_{ce}$) up to the Nyquist frequency of this study ($\sim$3906 Hz). First, broadband electrostatic waves observed at frequencies comparable to the ion plasma frequency ($f_{pi}$, in the plasma rest frame), which are linearly or elliptically polarized, and are parallel or oblique to the magnetic field are referred to as ion acoustic-like waves. Ion acoustic waves are thought to be generated by ion-ion or electron-ion drifts \citep{Gary1975,FG1984}, or by heat flux instabilities or nonlinear wave decay \citep{Dum1980,DO2006}. In the case where these waves show frequencies which are dispersive in time, they are denoted dispersive electrostatic waves (DEW), though the frequency drift is not definitely determined to be due to dispersion. ECDI-driven waves are characterized by having integer or half-integer electron cyclotron harmonics, ``comma''-shaped polarizations, and choppy waveforms \citep{W2010,Bren2013,W2014b}. They have been recently studied in shock ramps from simulations \citep{ML2006,ML2013,ML2017,MS2006}, which suggest the ECDI waves should occur in the foot and ramp. The polarizations and choppy waveforms are due to coupling between Bernstein waves and ion acoustic waves. ECDI waves are generated by the interaction between reflected ions and incident electrons \citep[][and references therein]{Forslund1972,W2010,Bren2013},
and typically have significant electric field components both parallel and perpendicular to the magnetic field. There are only a few published identifications so far of ECDI-driven waves near shocks \citep{W2010,Bren2013,W2014b,Goodrich2018}, though there is reason to believe some waves previously identified as IAW are actually ECDI-driven \citep{H2006}. Ion acoustic-like waves and Bernstein waves are sometimes observed at the same frequencies, and are difficult to distinguish (using automated search algorithms) without taking into account polarization differences.

Whistler mode waves, which are electromagnetic and right hand polarized, have been observed (in the solar wind) both upstream and downstream of IP shocks in two (not necessarily disjoint) frequency bands. The lower frequencies tend to be approximately equal to the lower hybrid frequency ( $f_{lh}\mathbf{\sim \sqrt{f_{ci}f_{ce}}}$ ) (like those frequently observed upstream of the Earth's bow shock \citep{Hoppe1982,HR1983,W2016}) and are usually upstream of IP shocks.
The higher frequency whistler waves have frequencies between the lower hybrid frequency and the electron cyclotron frequency, $f_{lh} < f_{whistler} \lesssim f_{ce}$, \citep{F1974,C1982,Bren2010,W2012,W2013,W2017,Giagkiozis2018,Goodrich2018,Cattell2020}, with common occurrence of frequencies around 0.15 to 0.3 $f_{ce}$ \citep{Bren2010,W2013,Cattell2020}.
%,Grul2017,Giagkiozis2018. 
They are large amplitude (sometimes dE$>$40 mV/m, dB $>$ 2 nT) and are oblique to the magnetic field with significant parallel electric fields. They are frequently observed downstream of IP shocks or in association with stream interaction regions.

\begin{figure}
\includegraphics[scale=0.45,keepaspectratio]{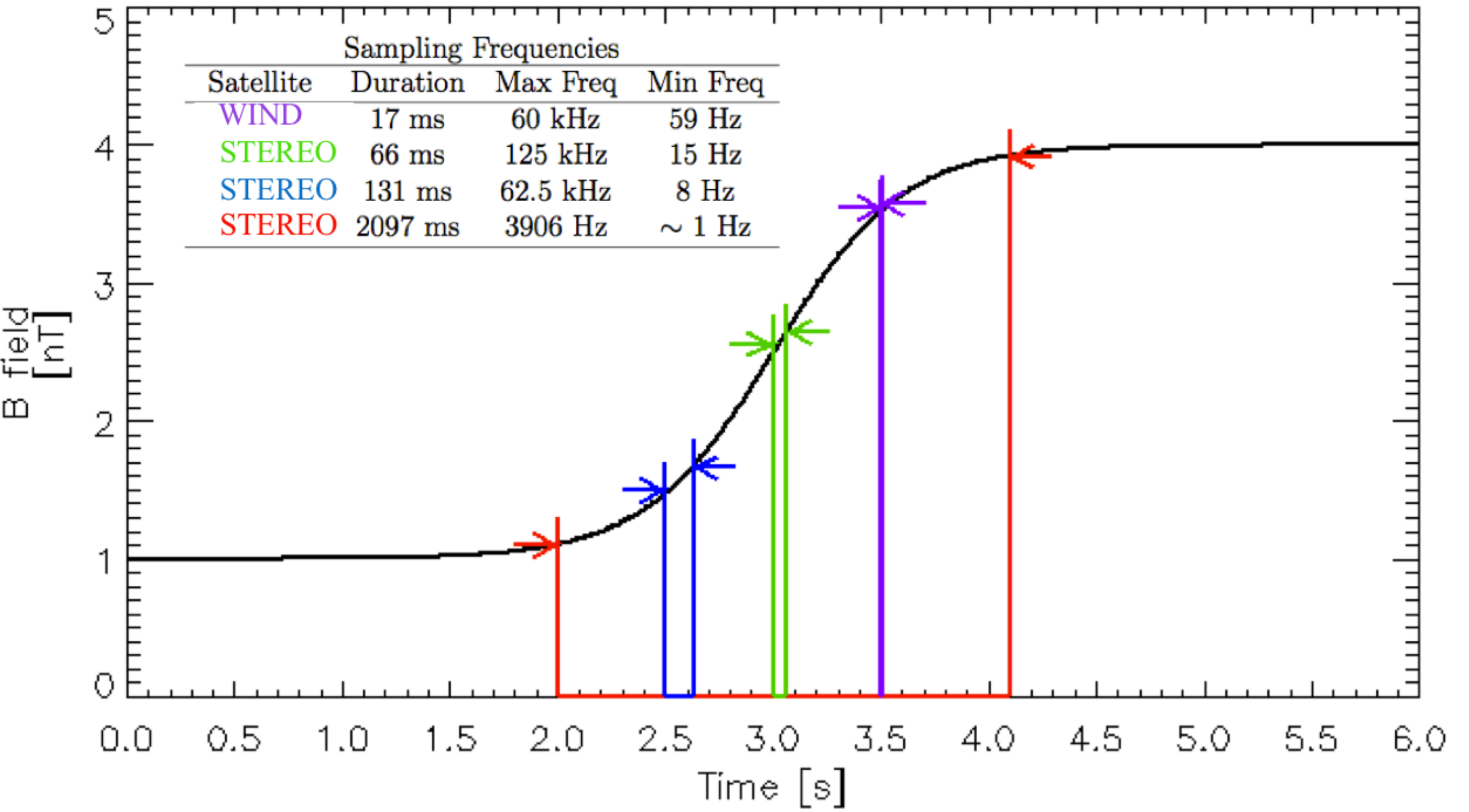}
\caption{The relative durations of the electric field measurement bursts on previous studies using STEREO and WIND. A mock shock ramp is plotted, with a duration comparable to what is seen in this (and other) studies, and full coverage can be attained using the $\sim$2s TDS mode of STEREO.}
\label{TDSlength}
\end{figure}

Previous studies have typically focused on waves in the ion acoustic (doppler shifted $\sim$ 1--10 kHz) to Langmuir (10s of kHz) regimes \citep[e.g.,][]{GA1977,G1979,H2006,W2007}, and on the lower frequency ($\lesssim$1 Hz) upstream (magnetosonic) whistler mode waves \citep{Hoppe1982,HR1983,RH1983,W2016}.
 This study focuses on plasma waves with frequencies from $\sim$10Hz--$\sim$4 kHz, to understand the importance of waves in this frequency range in the structure of shock ramps, as well as energy dissipation mechanisms associated with the shock ramps. This frequency range permits observation of some Doppler shifted ion acoustic-like waves ($\gtrsim$1 kHz), electrostatic waves showing frequency changes with time, ECDI-driven waves, and whistler mode waves (typically $\lesssim$ 60 Hz), as well as ESWs. Examples of each mode are shown in Figure \ref{exampleWaves}, described in detail in the next section. Furthermore, we utilize time domain waveforms over intervals more than 10 times as long as previous studies at IP shocks (Figure \ref{TDSlength}, described later in this section), allowing for observation of waves and their packet structure throughout the transition region of a shock in a single capture.  Direct observation of Langmuir mode waveforms and packets is not possible in this study since the electron plasma frequency is $\gtrsim$ 9 kHz. However, spectral data at lower time resolution can be used to observe the presence of possible Langmuir waves near the shocks. Note that we focus on waves with absolute maximum (zero-to-peak) amplitudes $>5$mV/m.

Figure \ref{TDSlength} demonstrates that this study has the ability to observe evolution of waves throughout the shock transition region and ramp. Previous studies using Wind with burst electric field captures of 17 ms and a lower frequency bound of $\sim$ 60 Hz could not observe wave evolution continuously in the ramp nor the high frequency whistler mode waves. Figure \ref{TDSlength} shows a simulated shock ramp, with a transition region of about 1-1.5 seconds, comparable to the average ramp duration of $\sim$1.2 seconds seen in this study. Longer bursts are vital for being able to observe the evolution and variability of waves throughout the ramp, as we show with the dramatic changes in wave modes and amplitudes that could not be studied with earlier instruments. Cluster, MMS, and THEMIS (and thus ARTEMIS) have had waveform captures long enough to cover the ramp regions of quasi-perpendicular bow shocks (as discussed in \citet[]{Balikhin2005,Kr2013,Goodrich2018,Goodrich2019} ). ARTEMIS has also had long waveform captures in interplanetary shocks. %perhaps add a citation to Lance's paper once it's on arXiv

In Section 2, we describe the instrumentation and methodology for identifying IP shocks and the wave modes observed in the time domain sample (TDS). In Section 3, we describe in detail the observations of two shocks to highlight the variability in the wave modes seen through the ramp and in the upstream and downstream regions. Section 4 describes the statistics for the 13 events. Discussion and the conclusions are presented in Section 5.

\section{Instrumentation and Methodology}
The STEREO satellites have heliocentric orbits slightly inside (STEREO-A) or outside (STEREO-B) 1 AU to respectively lead and lag Earth. Thus the spacecraft locations will vary in each of the time periods in this study. During the 2011 interval, the satellites were on opposite sides of the Sun, along a line approximately perpendicular to the earth-sun line. During the 2017-2018 interval, only STEREO-A was operational.

\subsection{STEREO Fields and Particles}
The STEREO WAVES (or S/WAVES) instrument \citep{Boug2008} measures the 3D electric field of waves in the solar wind. The time domain sampler (TDS) \citep{Boug2008,Bale2008} acquires 16384 samples per burst, and has four settings with sampling rates (capture durations) of 250 kilosamples/s (66 ms), 125 kilosamples/s (131 ms), 31250 samples/s (524 ms), and 7812.5 samples/s (2.097 s). Data exceeding a selection amplitude and onboard quality threshold is saved around the largest amplitudes and sent to the ground. These data will be referred to as ``TDS captures'' (or ``TDS samples''). The antennae on STEREO are responsive to density fluctuations in addition to the electric fields, but each antenna will be affected similarly to the others, and a pseudo-dipole channel can be included to isolate low frequency waves from density fluctuations on the scale of the STEREO spacecraft \citep{Bren2010,Cattell2020} . High frequency (2.6 kHz - 16.025 MHz) electric field intensities are measured with the  low frequency receiver (LFR) and high frequency receiver (HFR), which are averaged each minute and will be referred to as ``spectral data.'' All frequencies examined in this study are in the spacecraft frame.

We utilize data from three instruments in STEREO IMPACT \citep{Luhmann2008}. The 3D magnetic field is measured by the fluxgate magnetometer (MAG) instrument \citep{Acuna2008}, which has a normal mode of 8 samples/s and a burst mode of 32 samples/s. The solar wind electron analyzer (SWEA) \citep{S2008} provides electron distributions for energies from $\sim$50 eV up to 3 keV
and the suprathermal electron telescope (STE) \citep{L2008} measures electron flux for energies from 2 to 100 keV. The core of solar wind electrons is not observed.

The STEREO PLASTIC instrument suite \citep{Galvin2008} measures moments of proton distributions from energies of $\sim$0.3 to 80 keV/e. Bulk flow speed, density, and temperature are derived using a 1D Maxwellian fit of the moments, and the data are averaged over 1 minute.

Following solar conjunction, STEREO-A was rotated about the sun-spacecraft line by 180$^\circ$. As such, the view directions of many of the particle instruments are no longer along the nominal Parker spiral, but are rather perpendicular to it. Nonetheless, it is expected that this does not affect the observation of energetic particle events, but it does hinder the observation of beams streaming along the Parker spiral [private communication, \textit{R. Mewaldt, J. Luhmann, D. Larson}].

\subsection{Shock Identification}
This study focuses on the 2.097 s (from here referred to as 2.1s) duration TDS captures, taken by the S/WAVES instrument. During 2011, there were intermittent times from June-November when S/WAVES operated in the 2.1s burst mode. In March 2017, the instrument was switched to only take data in this 2.1s mode, and our study includes the interval from March 2017 - January 2018. During the first time period both STEREO-A and STEREO-B were operational; only STEREO-A was operational during the second. We limit our main observations to quasi-perpendicular shocks (angles to the upstream magnetic field $\theta_{Bn}$ $\ge 45^{\circ}$) where there are TDS samples within 1500 proton gyroradii ($\rho_{gi}$) of the shock ramp. This ranges from $\sim$60,000 km to $\sim$80,000 km for our events. For all shocks in this study this is the distance that has the majority of TDS captures which were stored by STEREO within a reasonable distance of the shock front.
The distance of 1500 $\rho_{gi}$ was chosen to preserve a consistent length scale, rather than a time scale, between shocks. %, which can have largely varying associated solar wind speeds.

Parameters for the shocks were obtained using the Rankine-Hugoniot relations along with upstream and downstream data for the magnetic field, density, bulk ion velocity, and ion temperature. We follow the same approach as that of \citet{KS2008} and \citet{VS1986} in solving the Rankine-Hugoniot relations to determine the local shock normal unit vector.  The asymptotic upstream/downstream time ranges were chosen based upon those times when the most parameters remained as stable as possible for a sufficient duration to have at least 10 data points on either side of the shock.  The time ranges were also examined for consistency with the asymptotic parameter values much further from the shock ramp, i.e., avoid fluctuations and/or changes close to the shock ramp.  The approach has been tested and used in several refereed publications \citep{Kanekal2016,W2014a,W2014b,W2016}. Plasma parameters determined with these methods and their uncertainties are given in \ref{ShockParamAll}. Shocks in this data set were identified from large variations observed in the magnetic field, from the IMPACT fluxgate magnetometer, which were accompanied by increases in density, proton temperature, and bulk flow speed as measured by PLASTIC. The ramp regions of the shocks are determined using the magnetic field, since the 8 Hz cadence, and in a few cases the 32 Hz burst cadence, provides the highest time resolution of the relevant plasma parameters. 

%There has been discussion of possible misidentification of several types of plasma discontinuities as shocks, including contact, tangential, and rotational discontinuities \citep{Hudson1970,Hudson1971}. Generally, \citet{Hudson1970,Hudson1971} finds that contact discontinuities are not expected to be observed near 1 AU, tangential discontinuities show magnetic field compression of less than 20\%, and rotational discontinuities only show compression up to 6\% and don't necessarily have magnetic field and density values correlated.
%Thus, to qualify as a quasi-perpendicular forward shock, the following criteria (consistent with, though less stringent than, the \citet{Kilpua2015} shock identification criteria) were required:
%\begin{eqnarray*}
%&\dfrac{B_{down}}{B_{up}} &\geq 1.2 \\
%&\dfrac{N_{down}}{N_{up}} &\geq 1.2 \\
%&\lvert V_{down} - V_{up}\rvert &\geq 10 ~km~s^{-1} \\
%&\theta_{Bn} &\geq 45^\circ
%\end{eqnarray*}
%within the margin of error associated with each measurement.

\subsection{Wave Mode Identification}
Wave modes are identified primarily via visual inspection of the data, including analysis of the highest power observed frequencies, two-dimensional polarizations (hodograms), and the physical characteristics of their waveforms (e.g., amplitude, ellipticity).
To facilitate wave mode identification, the three electric field components are rotated into a magnetic-field-aligned coordinate system (FAC), where the magnetic field direction is taken to be a piecewise step function over each 0.125s interval (the sample rate for the DC magnetic field). The magnetic field unit vector is defined as the parallel direction of the FAC system. The cross product of this direction with the spacecraft $x$ direction defines one perpendicular direction (referred to as $E_y$ or $E_{\perp2}$), and the cross product of $E_{\perp2}$ with $E_{||}$ completes the right hand system (referred to as $E_x$ or $E_{\perp1}$). The field aligned coordinate system allows determination of the highest amplitude components relative to the magnetic field direction, which can be a useful heuristic for identifying the specific wave modes that are present. When the magnetic field direction is very rapidly changing over the timescale of a wave packet, we use a minimum-variance coordinate system, which can then be compared to the average direction of the normal magnetic field, $\mathbf{B_n}$.

%Figure for examples of each wave mode
\begin{figure}
\includegraphics[scale=0.75,keepaspectratio]{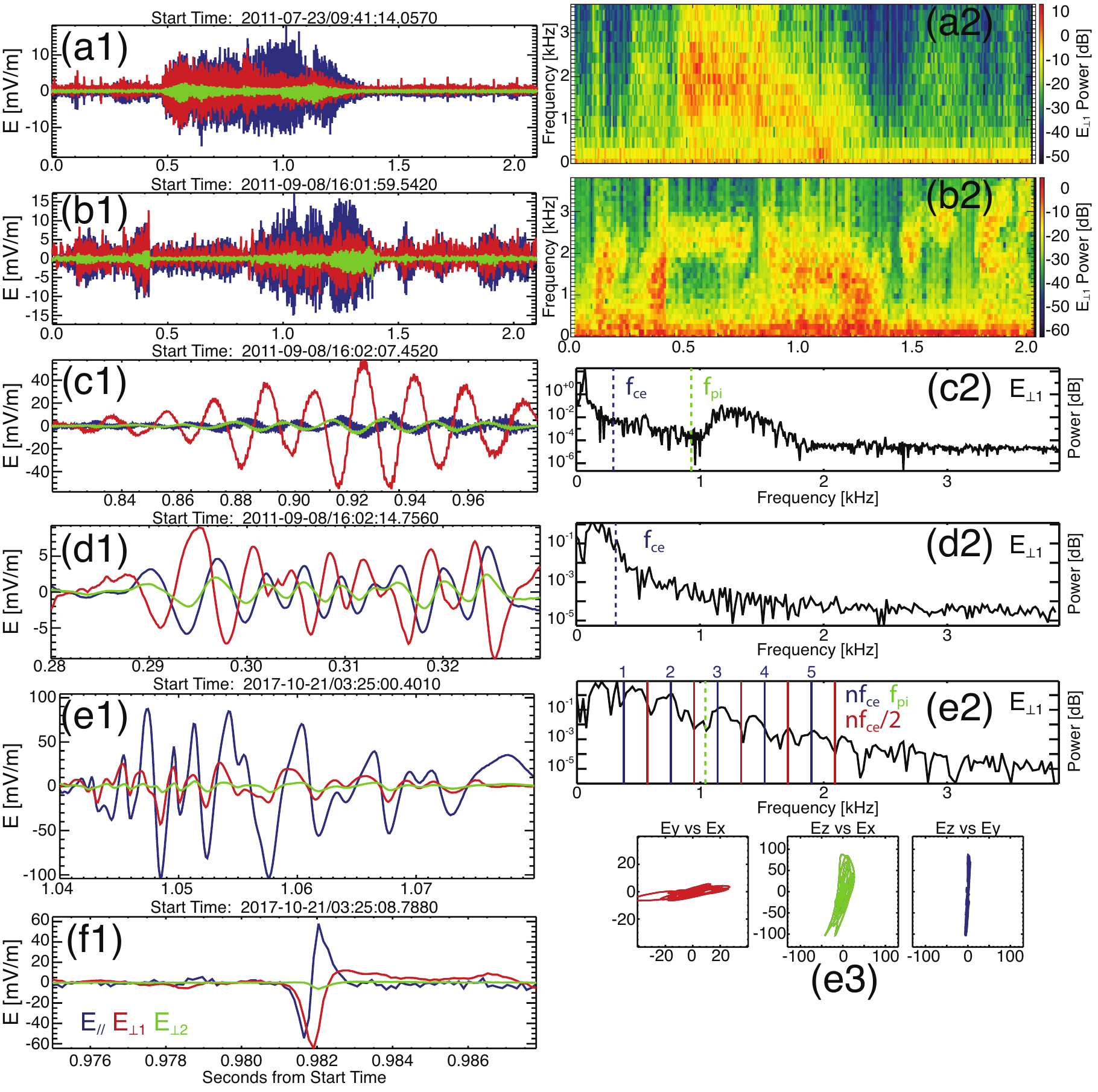}
\caption{Example TDS captures of waves observed in this study. (a1) An IA-like wave, with (a2) a sliding FFT on the right, showing broadband power in the kHz range. (b1) Several electrostatic packets showing frequency changes with time, with (b2) highest power frequencies changing over time from as low as a few hundred Hz up to a few kHz. (c1) A large amplitude, narrowband whistler mode wave. (c2) A large peak below the electron cyclotron frequency, as well as a broad peak above the ion plasma frequency suggesting a simultaneous IA-like wave. (d1) A broadband whistler mode wave, with (d2) power enhancement below and around the $f_{ce}$ (e1) An ECDI-like wave. (e2) FFT of the whole interval, with regular peaks at $f_{ce}$ harmonics and half harmonics. (e3) The hodograms for the ECDI-like wave are comma-shaped. (f1) An ESW, showing bipolar response parallel to the field and unipolar response in $E_{\perp1}$. In all panels, $E_{||}$ is in blue, $E_{\perp 1}$ is in red. Note time durations and amplitude ranges of samples differ.}
\label{exampleWaves}
\end{figure}

 Figure \ref{exampleWaves} presents examples of the wave modes categorized in this study; the durations of each plot are chosen to best showcase the wave being observed, and are not uniform. A wave packet is herein defined as an electric field waveform which has one point of at least 5 mV/m (zero-to-peak) in amplitude, and includes all adjacent points of at least 3.33 mV/m. Figure \ref{TDSDisp1105}, discussed later, shows examples of these packets. 
 Since the solar wind speeds are small compared to the speed of light $v/c << 1$, we know the contribution of any Lorentz-transformed magnetic field from the plasma frame to the spacecraft frame would be small compared to the typical E-field magnitude. The electric field parallel to the background magnetic field is shown in blue; one perpendicular component is shown in red. In Figure \ref{exampleWaves}a1-2, The wave begins as an ion acoustic-like (IA-like) wave, evidenced by a steady power spectrum sustained near the ion plasma frequency. The packet then displays frequency decreasing in time below the ion plasma frequency, then frequency increasing in time just before the packet ends, as shown in the sliding Fast Fourier Transform (sliding FFT) to the right. The main frequency component of the power spectrum goes from $\sim$ 1.4 kHz to $\sim$ 0.2 kHz in $\sim$ 0.25s. 
  As seen in Figure \ref{exampleWaves}b1-2, dispersive electrostatic wave (DEW) packets often cover a wide range of frequency space in the FFT (\ref{exampleWaves}b2), and are interspersed with IAW-like waves ($\sim0.5-0.8$s in  \ref{exampleWaves}b1-2). The change of frequency over time is not characteristic of IAWs, which are not dispersive, and so waves in a typical IAW frequency band which exhibit frequency changes in time (as in \ref{exampleWaves}a and in \ref{exampleWaves}b) are referred to with the shorthand dispersive electrostatic waves (DEWs). The change of frequency in time is not necessarily due to dispersion, and other possible mechanisms are discussed in sections 4 and 5. A clear whistler mode wave ($\sim$50 mV/m), Figure \ref{exampleWaves}c1-2, is observed with a simultaneous, much lower amplitude, ion acoustic wave (IAW). IAWs are identified by having a power spectrum near and above the ion plasma frequency, with linear or elliptical polarizations. The FFT over the whole time period (Figure \ref{exampleWaves}c2) shows that the whistler wave has a peak power below the local electron cyclotron frequency, $f_{ce}$ (determined with 8 Hz magnetic field data), and the IAW has power above the local ion plasma frequency, $f_{pi}$ (determined with 1 minute average density data). Figure \ref{exampleWaves}d shows a broadband whistler mode wave with less regularity, but still right hand polarized, which has power primarily below $f_{ce}$. Whistler mode waves have previously been identified in the spacecraft frame, as in \citet[]{Bren2010,Cattell2020}, and we follow the same methods of frequency and right-hand polarization analysis. In this example, regular peaks are seen in the power spectrum to the right, but these peaks are not spaced at $f_{ce}$ or its harmonics. Note that whistler mode waves typically have a dominant component in $E_{\perp}$ while IAWs tend to have a dominant $E_{||}$, though still with a large $E_{\perp}$. The wave in Figure \ref{exampleWaves}e shows the features described by \citet[]{Bren2013} to identify as generated by the ECDI. These include a choppy waveform (Figure \ref{exampleWaves}e1), peaks in the power spectrum corresponding to electron cyclotron harmonics and half harmonics (Figure \ref{exampleWaves}e2), and a comma-shaped hodogram for the parallel ($E_z$) and first perpendicular ($E_x$) components of the wave to the magnetic field (Figure \ref{exampleWaves}e3). Figure \ref{exampleWaves}f1 shows an electrostatic solitary wave, exemplified by the bipolar response parallel to the magnetic field and unipolar response in the perpendicular direction , consistent with previous research \citep[]{Andersson2009}.

Automated identification (auto-id) software was developed to assist in analysis of wave modes. 
In the auto-id program, TDS samples ($E_x,E_y,E_z,E_{dip}$) are discrete Fourier transformed to distinguish the highest power frequencies in several bands in user-selected time steps (typically from $\sim1-10$\% of the TDS) to observe the evolution of the dominant frequencies. For this study, the automated identification categorizes waves into 3 bands, 10 Hz -- 0.5$f_{ce}$, 0.5$f_{ce}$ -- $f_{ce}$, and $f_{ce}$--3906.25 Hz. 
Ion acoustic-like waves, dispersive electrostatic waves, and ECDI driven waves will sometimes be categorized in this study under the umbrella term of ``Intermediate Frequency Waves,'' where ``intermediate'' is referencing the range of frequency space lying between the electron gyrofrequency and the Nyquist frequency of this study. Since Langmuir waves cannot be directly observed in this study, we use the LFR/HFR spectral data to see whether there is significant power at or near $f_{pe}$ within a few minutes of the shock. In Figure \ref{exampleWaves}c, the auto-id would have been able to detect waves of intermediate frequencies and label them as ion acoustic-like, as well as the whistler mode wave and label it as such. However, due to the reliance on frequency alone, the software could not identify other wave modes that we are listing here.

\section{Observations of Waves at Two IP Shocks}

To demonstrate more specifically that wave modes associated with shocks are highly variable, both in and around the ramp region, we will show two cases: the highest Mach number shock in our set of events ($M_{f} = 5.4 \pm 1.0, \beta \sim 1.05, \theta_{Bn} = 77^\circ \pm 9^\circ$), on 2017-10-21, and the lowest Mach number shock ($M_{f} = 1.0 \pm 0.5, \beta \sim 0.17, \theta_{Bn} = 66^\circ \pm 8^\circ$), on 2011-11-05. For all shocks in this study, the TDS captures which were within 1500 proton gyroradii (computed based upon thermal velocity, $v_{th}$) of the shock front were analyzed. (For average solar wind speeds, 1500 $\rho_{gi}$ $\sim$87300 km upstream, $\sim$ 63000 km downstream; or $\sim$ 254s upstream, $\sim$ 166s downstream; however, each event was analyzed using the in situ measurements of plasma parameters, e.g., solar wind speed, not the average.)

\begin{figure}[htb!]
\centering
%\hspace{12pt}
\textbf{Event Overview 2017-10-21}
\includegraphics[width=6in,keepaspectratio]{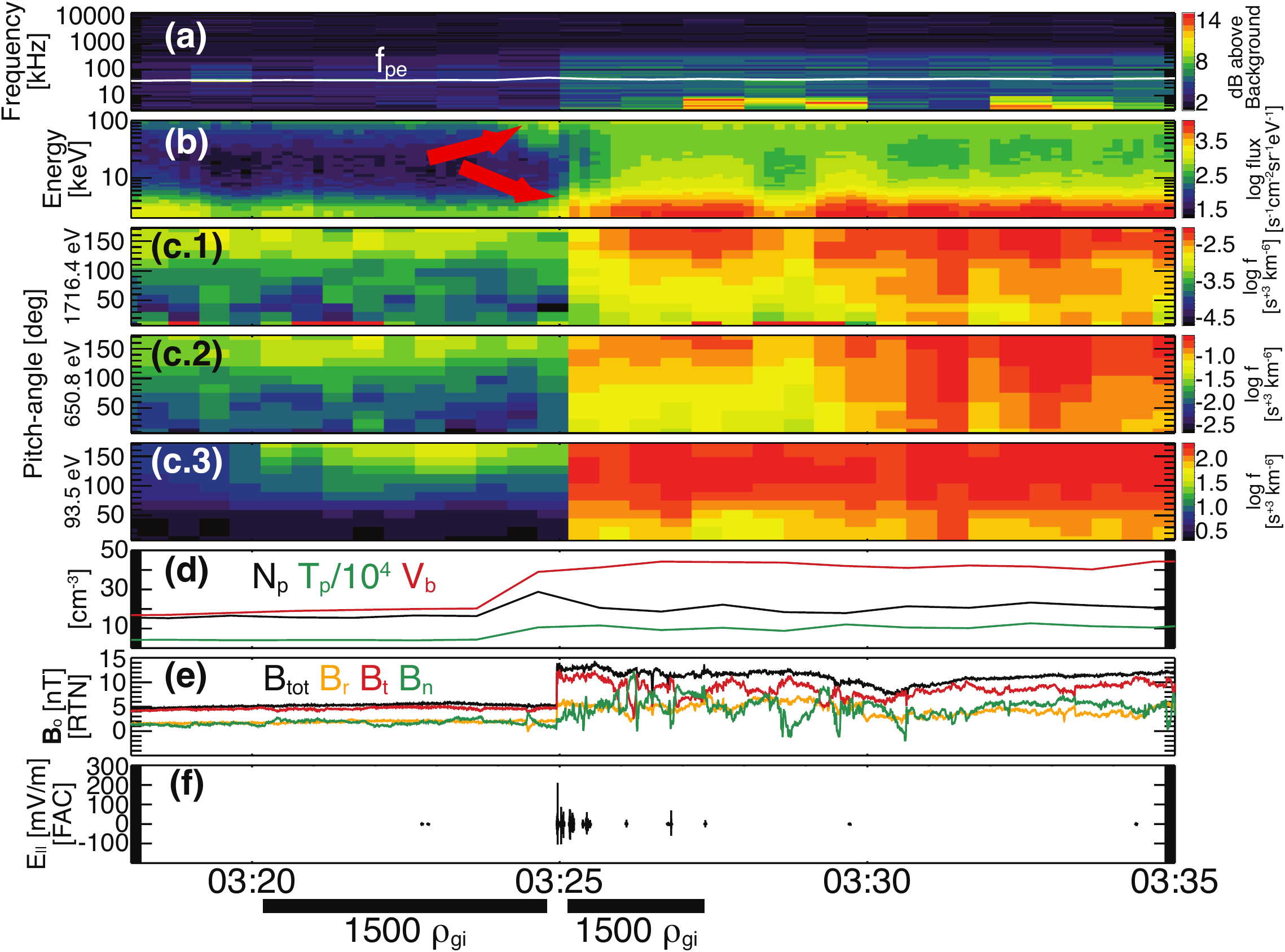}
\caption{Shock event on STEREO-A on 2017-10-21. (a) Electric field intensity above background averaged over each minute around the shock. The electron plasma frequency is overlain. (b) Electron number flux in energies 2-100 keV over time from STE, with red arrows denoting energy enhancements upstream. (c1-3) Electron pitch angle distributions from SWEA, at the energies 1716.4 eV, 650.7 eV, and 93.5 eV, showing a peak at 180$^\circ$ showing electrons flowing anti-sunward. Color axis is distribution function in $s^3km^{-6}$. (d) The proton density (black, in cm$^{-3}$), temperature (green, in K), and velocity (red, arbitrary units), at a rate of once per minute. (e) The total magnetic field (black) and 3 components in RTN (orange, red, green, respectively). The data rate displayed here is 8 Hz. (f) The parallel component of TDS burst data from S/WAVES in the 2.1s mode. Braces denote 1500 $\rho_{gi}$ upstream and downstream of the shock.}
\label{1021Overview}
\end{figure}

An overview of the highest Mach number ($\simeq$5.4) shock in this study is shown in Figure \ref{1021Overview}.
The shock ramp occurs at $\sim$03:25, as evidenced in the jump in the magnetic field (Figure \ref{1021Overview}e) with the increase in bulk flow speed, density, and temperature (Figure \ref{1021Overview}d). Immediately following the shock, the 2-100 keV electrons in the downstream STE data show enhancements, in particular at the lowest energies (Figure \ref{1021Overview}b). We can infer these electrons are streaming away from the sun because the highest energy SWEA distributions (Figure \ref{1021Overview}c1) peaks near 180$^\circ$, and the radial magnetic field is toward the Sun (panel e). There is also a structure at high energy just upstream, and at lower energies upstream which suggest reflected particles (denoted with red arrows in Figure \ref{1021Overview}b). The higher energy electrons may be due to reflections of electrons off of the shock, but the necessary pitch angle distributions are not available to confirm this. There are also enhancements across the lower energies ($\sim$50 eV -- 1.7 keV) in the SWEA data (Figure \ref{1021Overview}c1-3) primarily with pitch angles distributed from 0$^\circ$ to 90$^\circ$ relative to the magnetic field, which here has a radial component ($B_r > 0$) pointing sunward, but with much broader pitch angles than upstream. The largest amplitude waves are observed during or very close to the ramp and in the region immediately downstream (Figure \ref{1021Overview}f). The spectral data (Figure \ref{1021Overview}a) shows enhanced power near the local plasma frequency for several minutes upstream of the shock, as well as possible enhancement a few minutes upstream, suggesting the presence of Langmuir waves (cf. \citet[]{W2007}, which observed Langmuir waves commonly upstream of shocks, which are more difficult to observe in this study).

\begin{figure}[htb!]
\centering
\textbf{TDS at the Ramp and Downstream}
\includegraphics[scale=0.9,keepaspectratio]{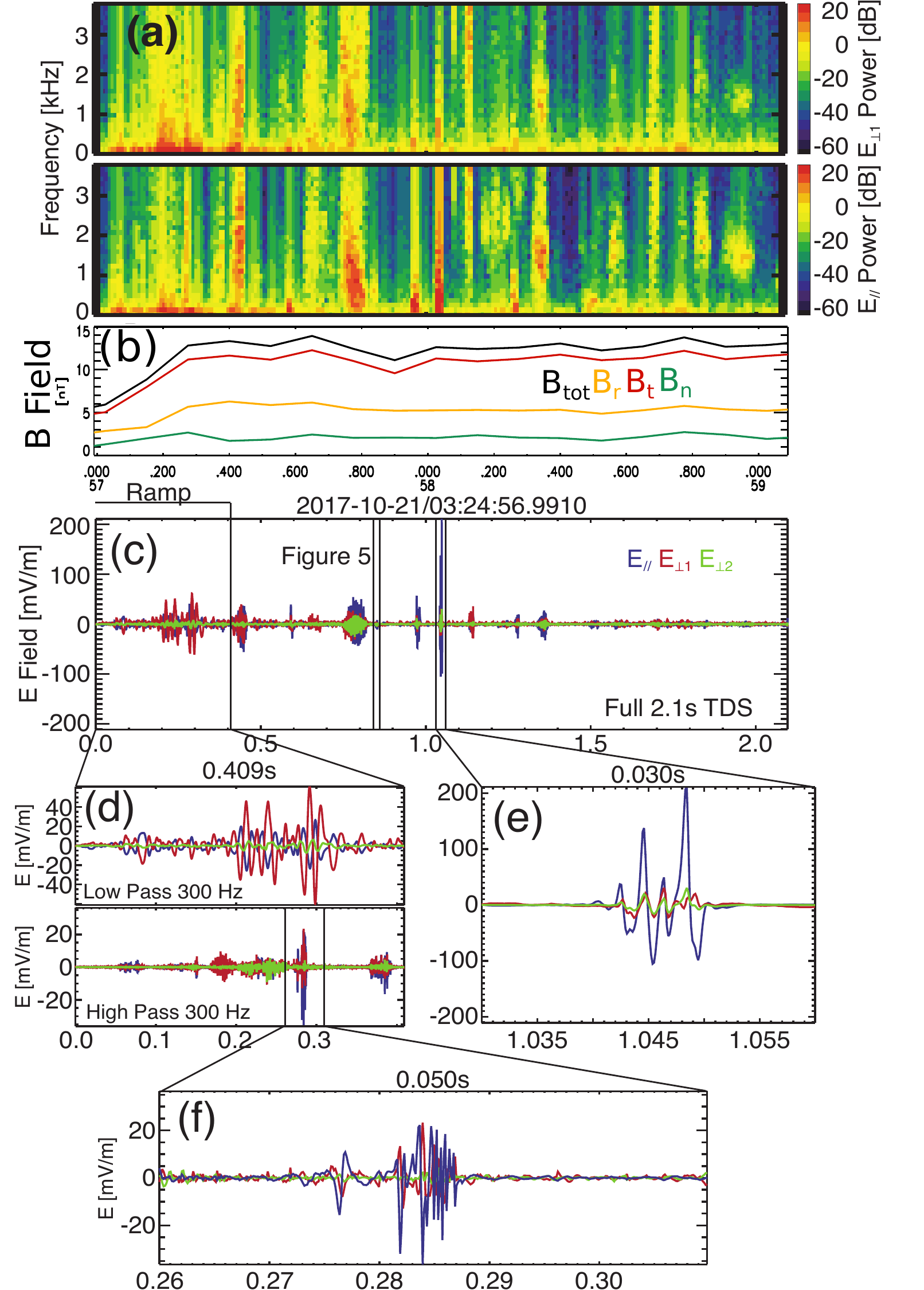}
\caption{TDS which overlaps with 2017-10-21 shock ramp. (a) FFT of one perpendicular (top, Ex) and one parallel component (second, Ez) of TDS over each 64 points. (b) Magnetic field components at 8 Hz.(c) Three components of the electric field near a shock on 2017-10-21, displaying the full 2.1s TDS. The shock ramp coincides with the first 0.409s of this TDS. The electric field is in FA coordinates. (d) Low and high pass filtered ramp waves. A whistler mode wave is seen in the upper diagram. (e) The largest amplitude wave in this TDS, an intermediate frequency wave of $\sim$200 mV/m in amplitude. (f) An ESW followed by an intermediate frequency packet in the ramp.}
\label{shock1021Ramp}
\end{figure}

The TDS capture shown in Figure \ref{shock1021Ramp} overlaps with the ramp region for the first 0.409 s, as seen by the magnetic field in \ref{shock1021Ramp}b. The electric field amplitude in the component parallel to the magnetic field (\ref{shock1021Ramp}c,\ref{shock1021Ramp}e) peaks at $>$200 mV/m. Furthermore, in the shock ramp itself we find whistler mode waves (\ref{shock1021Ramp}c,\ref{shock1021Ramp}d), intermediate frequency wave packets (including \ref{shock1021Ramp}c,\ref{shock1021Ramp}d,\ref{shock1021Ramp}e,\ref{shock1021Ramp}f), and an ESW (\ref{shock1021Ramp}c,\ref{shock1021Ramp}f), demonstrating that a shock ramp can contain several wave modes across a wide range of frequency space. For instance, during the ramp we see whistler mode waves occurring simultaneously with IAW-like waves. Additionally, there are intermediate frequency and electrostatic solitary packets throughout the downstream portion of this TDS. The highest amplitude wave packets in this capture are highly nonlinear and do not seem to coincide with the 
electron cyclotron harmonics or ion plasma frequency (\ref{shock1021Ramp}c,\ref{shock1021Ramp}e).

\begin{figure}[htb!]
\centering
\textbf{Event Overview 2011-11-05}
\includegraphics[scale=0.75,keepaspectratio]{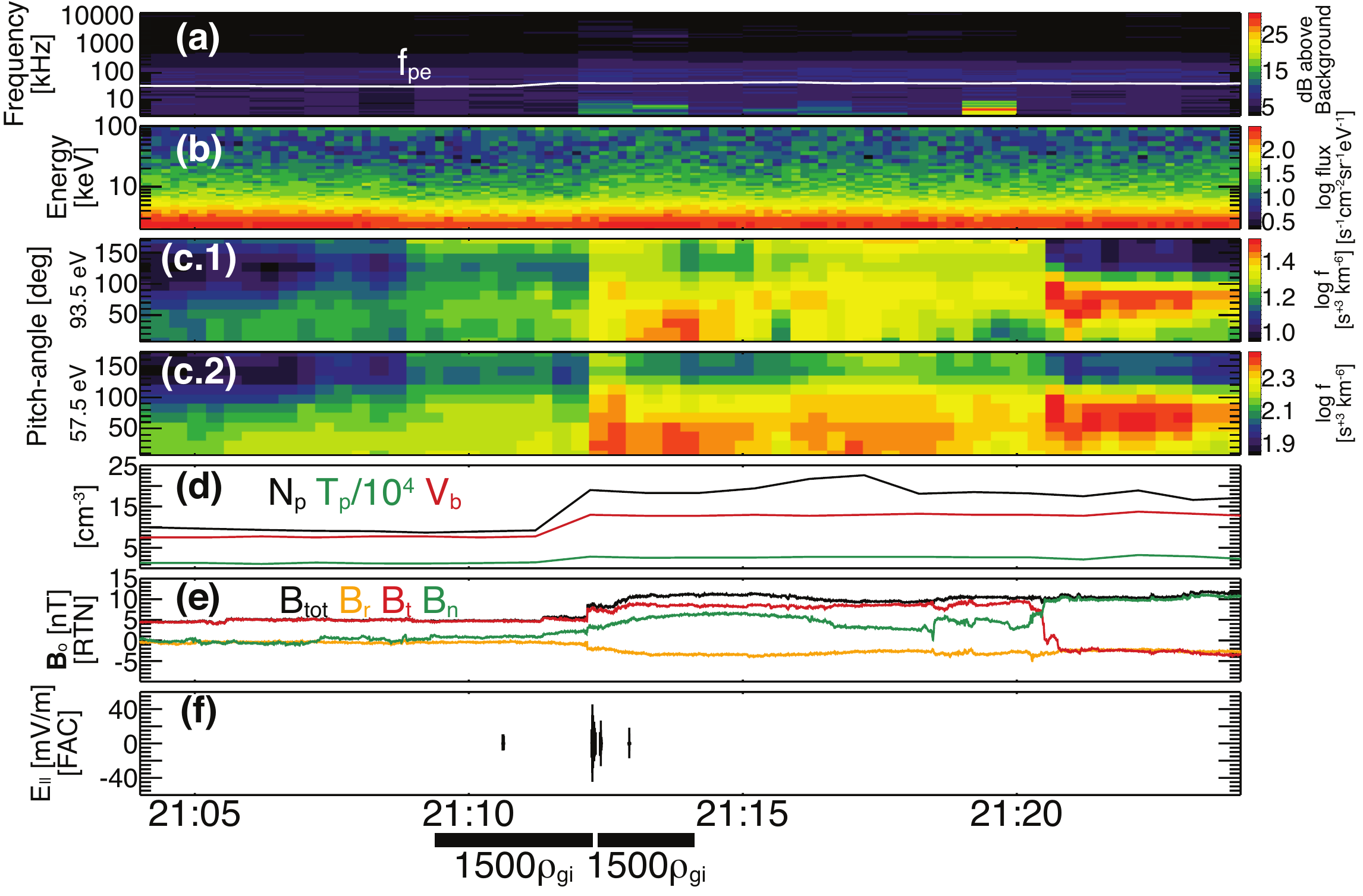}
\caption{STEREO-A with interplanetary shock on 2011-11-05. (a) Electric field intensity above background, averaged (1 minute resolution). (b) Electron counts in energies 2-100 keV over time from STE.  (c.1-3) Electron pitch angle distributions from SWEA, at the energies 93.5 eV and 57.5 eV. Color axis is distribution function in $s^3km^{-6}$. (d) The proton density (black, in cm$^{-3}$), temperature (green, in $10^4$K), and velocity (red, arbitrary units), at a rate of once per minute. (e) The total magnetic field (black) and 3 components in RTN (orange, red, green, respectively), at a rate of 8 Hz. (f) The parallel component of TDS burst data from S/WAVES in the 2.1s duration mode. Braces denote 1500 $\rho_{gi}$ upstream and downstream of the shock. Note that the color bars for STE and SWEA data are different in this figure compared to Figure \ref{1021Overview} due to lower particle fluxes}
\label{1105Overview}
\end{figure}

The event on 2011-11-05 provides further insight into the variability of waves within 1500 proton gyroradii of the ramp. Table \ref{ShockParamAll} (discussed in detail in section \ref{SummarySection}) shows that the Mach number for this shock is the lowest in our data set. The shock ramp occurs near 21:12, as seen in the proton data (Figure \ref{1105Overview}d) and the magnetic field data (Figure \ref{1105Overview}e). Note that the ramp structure is complex and the magnetic field indicates a clear `foot' structure upstream. The time resolution of the plasma data (Figure \ref{1105Overview}d) is not adequate to resolve this structure. The spectral data (Figure \ref{1105Overview}a) shows no clear enhancement near the plasma frequency upstream ($f_{pe}$$\sim$$27$kHz), in the ramp, or downstream ($f_{pe}$$\sim$$39$kHz), thus there is no evidence for large amplitude Langmuir waves. The SWEA data (Figure \ref{1105Overview}c1-2) show complex and changing enhancements to the electron distribution, though only in the lowest energy channels. There is a foot structure observed upstream, and there are enhancements in the perpendicular electrons slightly further upstream. The STE data show no clear enhancement in the $2$--$100$ keV electrons. The highest amplitude waves which have been transmitted are again immediately downstream (Figure \ref{1105Overview}f). As we discuss in more detail below, the waves upstream of this shock display frequency drifting characteristics, while the waves downstream have many signatures of being ECDI driven.

\begin{figure}[htb!]
\hspace{0.08in}\title{}
\centering
\includegraphics[height=300pt,width=300pt,keepaspectratio]{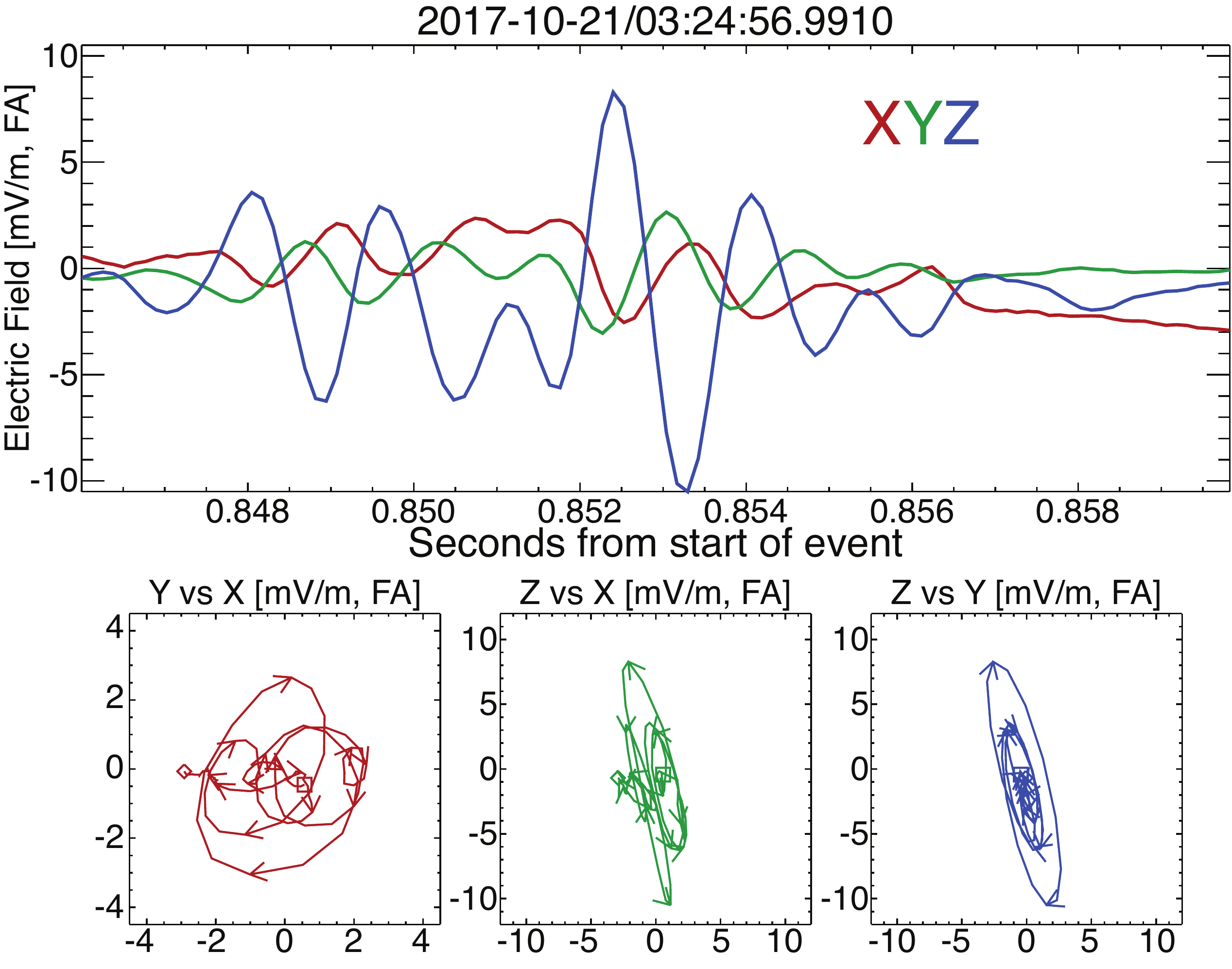}
\caption{(Top Panel) Three components of the electric field in FAC. (Bottom) Hodograms of each pair of components. The hodograms evidence an elliptical, left-hand polarized wave. A square denotes the start of each hodogram, and a diamond denotes the end. Arrows trace the direction the hodogram travels in time. The peak frequencies are near $\sim$450 and $\sim$650 Hz, with a local $f_{ce} \sim$350 Hz and a local $f_{pi} \sim$950 Hz.} 
\label{leftHand1021}
\end{figure}

\begin{figure}[htb!]
\hspace{0.08in}\title{TDS Upstream of Shock}
\centering
\includegraphics[height=400pt,width=400pt,keepaspectratio]{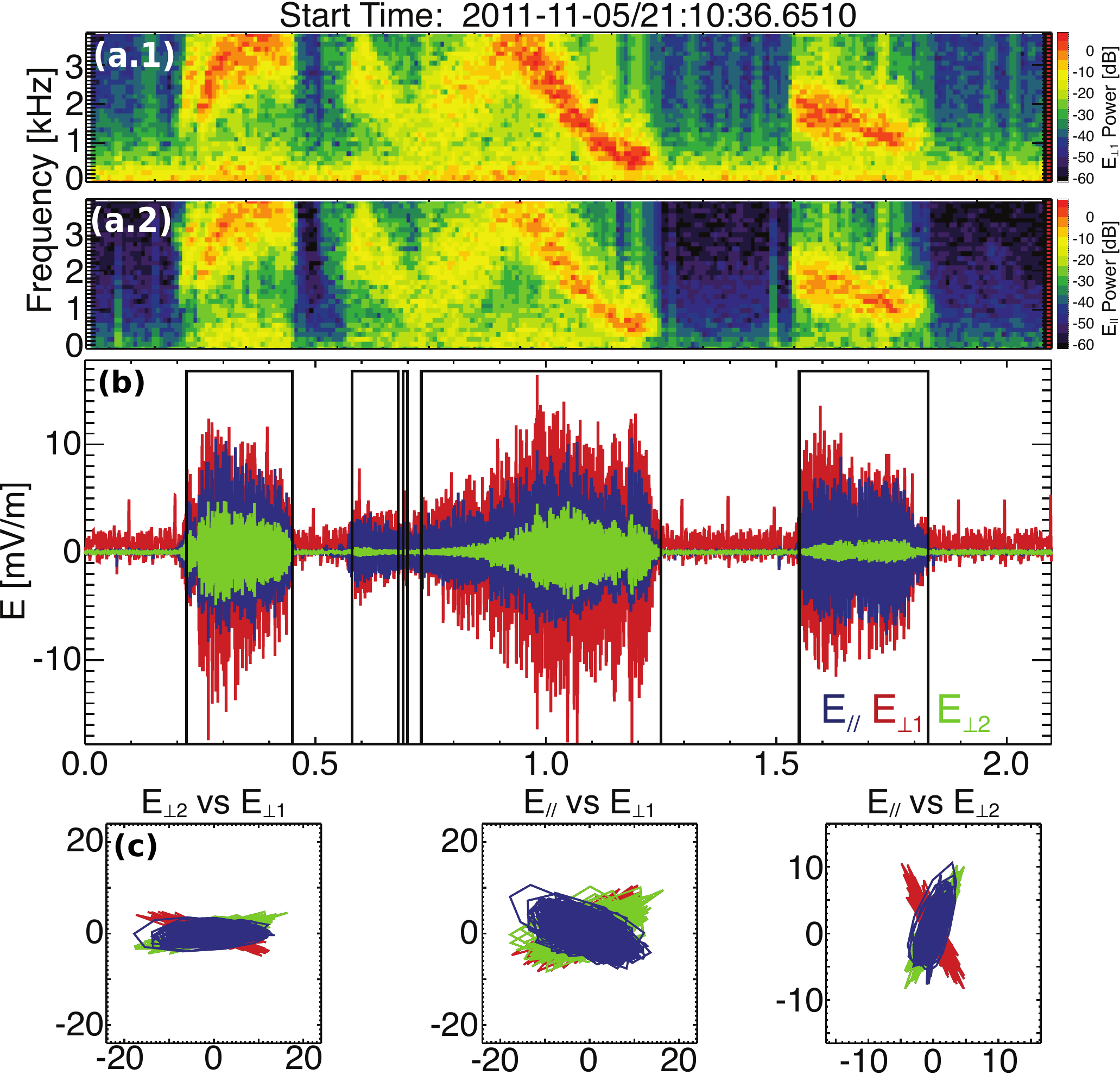}
\caption{STEREO-A TDS upstream of 2011-11-05 shock event. (a.1-2) FFTs of one perpendicular ($E_x$) and the parallel component ($E_z$) of the electric field. The ion plasma frequency is plotted as a black line. (b) The parallel and both perpendicular components of the electric field in a 2.1s TDS. Wave packets are shown marked by black outlines, with the vertical lines falling at the beginning and end of each packet. (c) Three hodograms showing polarization cross sections over the whole time interval in field aligned coordinates. The polarization changes in time are denoted by blue, then green, then red in time.}
\label{TDSDisp1105}
\end{figure}

There are several waves of special interest encountered throughout this study, as they prove difficult to classify as one of the usual modes identified in previous studies. Upstream of the 2011-11-05 shock, we see waves with significant frequency drift and polarization changes occurring in concert with change in the sign of the drift (Figure \ref{TDSDisp1105}, discussed below). Downstream of this shock, there are a few ECDI-like packets, the final of which grows in amplitude and shifts in frequency into what appears to be an IA-like power spectrum. Near 0.4 seconds downstream of the 2017-10-21 shock, in Figure \ref{leftHand1021}, we see the hodograms evidence an elliptically polarized wave. The FFTs show the peak frequencies are near $\sim$450 and $\sim$650 Hz, with a local electron gyrofrequency of $\sim$350 Hz and a local ion plasma frequency of $\sim$950 Hz. The wave has been rotated to the nearest local magnetic field measurement. The leftmost hodogram shows the trajectory of the wave around that magnetic field line, which comes out of the page. We can see that the wave generally travels in a left-handed manner relative to the most recent magnetic field measurement. Note that because we do not know the mode of this wave, nor whether it is electrostatic or electromagnetic, we cannot determine the wave vector $\vec{k}$. It is worth noting that the magnetic field may have gone through a rotation since the previous magnetic field measurement. However, the magnetic field measurement immediately preceding this and the magnetic field measurement immediately following this differ by less than 5$^\circ$. More observations could provide better insight to this phenomenon, and could clarify whether this is in fact a previously unobserved mode.
Two wave packets further downstream of this shock appear highly nonlinear, show evidence of electron cyclotron harmonics, consistent with ECDI, and have amplitudes of $\sim$60 mV/m and $\sim$200 mV/m.

Upstream of the 2011-11-05 shock, we observe a TDS which displays large changes in the frequency during each of three separate wave packets (Figure \ref{TDSDisp1105}). The FFTs of one perpendicular ($E_x$) and one parallel ($E_z$) component are plotted (Figure \ref{TDSDisp1105}a1-2), demonstrating that the frequencies of packets can increase, decrease, or display behavior that drifts both positively and negatively. The frequencies observed here are primarily above the local ion plasma frequency ($f_{pi}$). The waveforms are shown (Figure \ref{TDSDisp1105}b) with packets outlined by black lines, denoting where the waves are at least 3.33 mV/m in amplitude, with nearby amplitudes of at least 5 mV/m. Furthermore, from the hodograms, we see that the polarization of the wave packets is not consistent throughout the TDS (Figure \ref{TDSDisp1105}c), switching direction during and at the end of the second wave packet. The polarization in Figure \ref{TDSDisp1105}c from 0s to 0.8s is shown in blue, the polarization from 0.8s to 1.1s in green, and the polarization from 1.1s to 2.1s in red. Note that these times approximately coincide with changes in the sign of the derivative of the frequency change, suggesting that the wave packets may be propagating in different directions. The lowest frequencies in the second packet are $\sim f_{pi} \approx 630$ Hz. A Fourier transform of the whole time range reveals separated peaks, but they are not clearly spaced either at $f_{ce}$ nor at $0.5 f_{ce}$, thus we classify this as a ``dispersive" electrostatic wave. The mechanisms for DEWs like this have not yet been determined, though some possibilities are described in sections 4 and 5.

\begin{figure}[htb!]
\hspace{0.08in}\title{TDS Downstream of Shock}
\centering
\includegraphics[height=400pt,width=400pt,keepaspectratio]{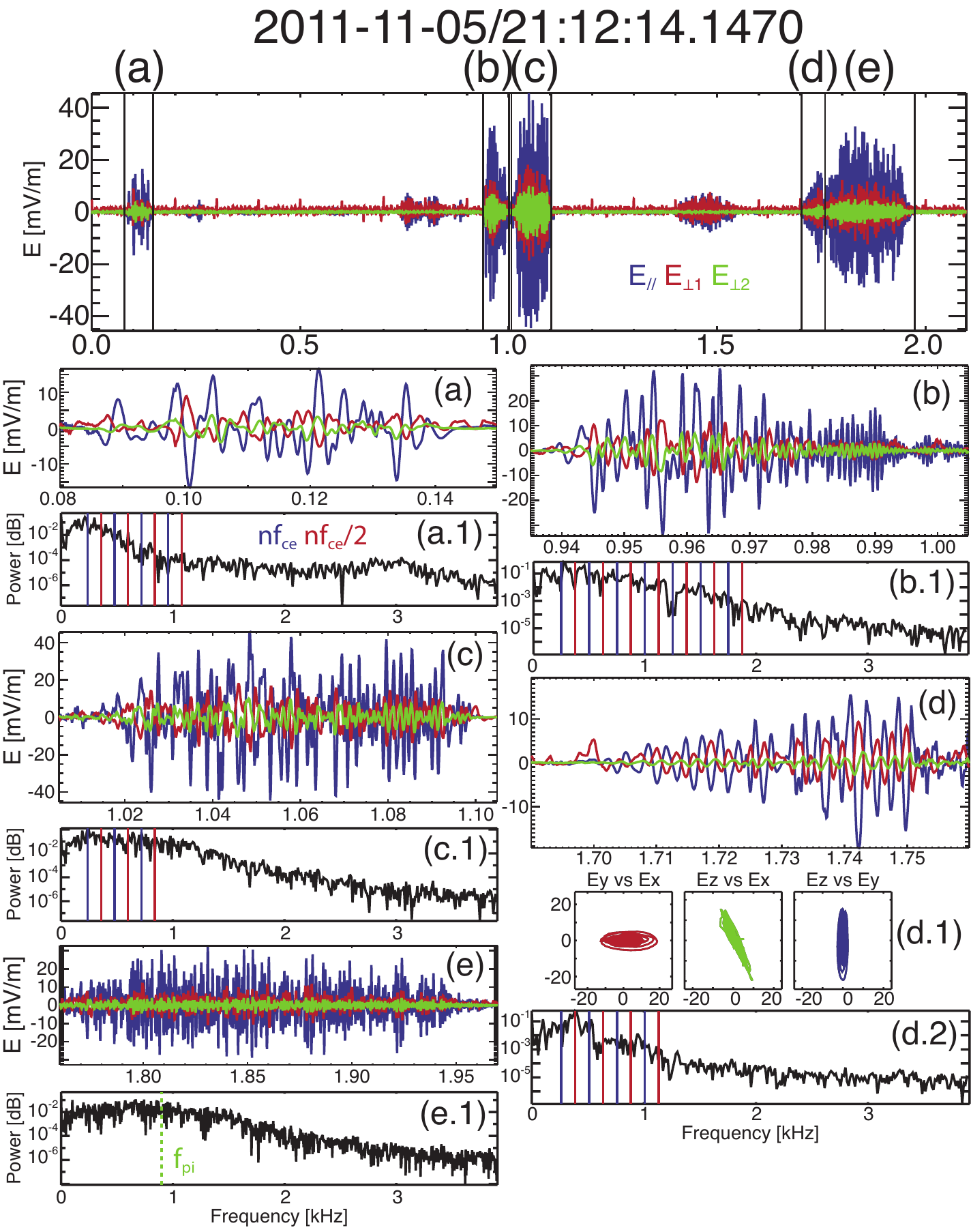}
\caption{STEREO-A TDS downstream of 2011-11-05 shock event. (Top) Full 2.1 second TDS, with several packets and an instrumental effect in $E_{\perp1}$. (a) An ECDI-like wave and its FFT, showing peaks spaced at about 0.5$f_{ce}$ which do not always coincide with the precise harmonics. (b) An ECDI-like wave with large peaks near $f_{ce}$ harmonics and half harmonics. (c) An ECDI-like wave with broad power peaks near $f_{ce}$ harmonics. (d) An ECDI like wave with elliptical and comma shaped hodograms. FFT peaks coincide with harmonics at 1.5,2,3, and 4 $f_{ce}$. (e) An IAW-like wave with broad power around $f_{pi}$. Note that the top panel indicates packet durations.}
\label{TDSDownstream1105}
\end{figure}

In contrast, as shown in Figure \ref{TDSDownstream1105}, a number of the waves downstream of the 2011-11-05 shock demonstrate regular peaks, coinciding with integer and half-integer harmonics of the electron cyclotron frequency. Furthermore, the broad frequency peaks are either greater than or coincide with the local ion plasma frequency. Note that Doppler shift has not been taken directly into account. The hodograms are not strongly comma-shaped, but do exhibit irregular polarizations, which are neither consistently elliptical nor consistently linear. Thus the majority of these wave packets have been identified as ECDI-driven waves, with the remainder classified as ion acoustic-like waves.

The 5 TDS captures which overlapped with the ramps of interplanetary shocks showed that every wave mode that could be observed within the constraints set by the Nyquist frequency was present in at least one shock ramp. Notably, there was a capture overlapping the end of a shock ramp (with relatively high $\beta$ for this study) which showed no large amplitude waves within the observed frequency range for that duration of the ramp. While the TDS captures did not have perfect coverage of all ramps, it is interesting to note that the observed waves are rapidly varying within the very short duration of the shock ramp.

\section{Summary of Shocks with TDS}\label{SummarySection}

A total of 12 shocks from the 2011 and 2017-2018 intervals had at least one 2.1s TDS burst within 1500 proton gyroradii (for average solar wind speeds, $\sim$87300 km upstream or $\sim$ 63000 km downstream, or on the order of $\sim$10-15 $R_E$) of the shock ramp. The shock parameters are shown in Table \ref{ShockParamAll}. The downstream (subscript d) to upstream (subscript u) ratios for magnetic field magnitude, density, and temperature are average values over a period of 5 minutes near the shock. The fast Mach numbers ranged from 1.0 to 5.4. Note that the Mach numbers and shock normal angles to the magnetic field are difficult to determine, and are all consistent within their margins of error with quasiperpendicular shocks (Table \ref{ShockParamAll}). An estimate of the first critical Mach number, obtained from results in \citet{EK1984}, indicates that 3 of the shocks are supercritical, with $M_f$/$M_{cr}$ ranging from 0.2 to 2.0.
Of these, 5 shocks had at least partial TDS coverage of the ramp (denoted by an asterisk on the date). The average ramp duration was $\sim$1.2s, with an average coverage of $\sim$70\% of the ramp for the five cases. Four of the five ramp TDS had waves with amplitudes larger than 5 mV/m, while the fifth had no waves $>$5 mV/m within the observed portion of the transition region.
\FloatBarrier
\begin{table}
\footnotesize
\caption{Shock parameters of 12 forward shocks in which we had TDS captures within 1500$\rho_{gi}$ of shock ramp. The dates with asterisks (*) denote the shocks for which there were TDS captures partially or completely overlapping the shock ramp. Underlined dates had magnetosonic whistler precursors. Daggers (${\dagger}$) indicate supercritical shocks. Velocities are in km/s. Magnetic field data are from IMPACT; N,T, and V data are from PLASTIC proton measurements. Subscript u(d) represents the upstream(downstream) value. The number of TDS in the upstream (downstream) region is given by $\#U$(D), with peak amplitudes in mV/m. The quantity $\lvert3k_B\Delta T_p / \Delta K\rvert$ represents the change in proton thermal energy compared to the change in bulk flow kinetic energy across the shock, as computed in the normal incidence frame, with delta representing the change from the upstream to the downstream parameters.} 
\centering
\tabcolsep=0.11cm
\begin{tabular}{l c c c c c c c c c c c c}
\hline
\ \ \ \ Date & Time & $B_{d}/B_{u}$ & $N_{d}/N_{u}$ & $\lvert3k_B\Delta T_p/\Delta K\rvert$& $\lvert{\Delta V}\rvert$ & $\beta$ & $M_{f}$ & $\theta_{Bn}$ & $V_{sh}$ & $M_f/M_{cr}$ & $\#U$(pk) & $\#D$(pk)\\
\hline
\underline{2011-07-23} & 09:41:10 & $1.78 \pm {0.16}$ & $3.19 \pm {0.39}$ & ${0.81} \pm {0.74}$ & 30$\pm {6}$ & 0.94 & 1.3$\pm$0.3 & $48^\circ\pm10^\circ$ & 390$\pm {34}$ & 0.5$\pm$0.1 & 1 (20.95) & 3 (18.32) \\ % 0.8/1.2
\underline{2011-08-06} & 12:42:39 & $2.64 \pm {0.22}$ & $3.70 \pm {0.54}$ & ${0.16} \pm {0.12}$ & 57$\pm {16}$& 0.65 & 1.8$\pm$0.4 & $73^\circ\pm14^\circ$ & 421$\pm {56} $& 0.8$\pm$0.2 & 0 & 1 (9.75) \\ % 1.4/1.2
2011-09-08 & 16:01:27 & $3.71 \pm {0.31}$ & $2.27 \pm {0.32}$ &$ {0.29} \pm {0.17} $& 58$\pm {5}$ & 0.22 & 2.2$\pm$0.4 & $80^\circ\pm8^\circ$ & 370$\pm {19}$ & 0.9$\pm$0.1 & 0 & 14 (58.18) \\ % 1.4/1.3
\underline{2011-09-19} & 13:35:47 & $1.49 \pm {0.23}$ & $1.72 \pm {0.25}$ &$ {0.20} \pm {0.29}$ & 27$\pm {3}$ & 0.57 & 1.7$\pm$0.6 & 59$^\circ\pm9^\circ$ & 318$\pm {46}$ & 0.7$\pm$0.2 & 0 & 2 (5.8) \\ % 1.1/1.3
\underline{2011-11-05} & 21:12:09 & $2.28 \pm {0.17}$ & $1.91 \pm {0.34}$ & ${0.79} \pm {3.3} $& 24$\pm {2}$ & 0.17 & 1.0$\pm$0.5 & 66$^\circ\pm8^\circ$ & 259$\pm {32}$ & 0.4$\pm$0.2 & 1 (19.54) & 4 (49.25) \\ % 0.7/1.3
\hline
2017-05-04 & 21:01:27 & $1.39 \pm {0.28}$& $1.57 \pm {0.32}$ & ${0.16} \pm {1.5}$ & 26$\pm {3}$ & 0.51 & 2.1$\pm$2.7 & 60$^\circ\pm11^\circ$ & 261$\pm {141}$ & 0.9$\pm$1.1 & 0 & 1 (75.36)\\ % 2.7/1.2
\underline{2017-05-09}*$^{\dagger}$ & 10:30:39 & $2.66 \pm {0.16}$ & $4.17 \pm {0.47}$ &$ {0.63} \pm {0.24}$ & 96$\pm {6}$& 0.85 & 2.6$\pm$0.4 & 60$^\circ\pm14^\circ$ & 465$\pm {36}$ & 1.1$\pm$0.2 & 2 (50.86) & 2 (35.65) \\ % 1.7/1.2
2017-05-22*$^{\dagger}$ & 17:22:19 & ${1.46} \pm {0.34}$ & $1.61 \pm {0.09}$ &$ {0.14} \pm {0.10}$ & 24$\pm {3}$& 1.81 & 2.7$\pm$0.5 & 37$^\circ\pm14^\circ$ & {340}$\pm {30}$ & 1.5$\pm$0.3 & 0 & 6 (15.7) \\ % 0.9/1.1
\underline{2017-06-20} & 08:38:00 & $1.32 \pm {0.19}$ & $1.87 \pm {0.18}$ & ${0.13} \pm {0.16}$ & 22$\pm {3}$ & 1.40 & 1.6$\pm$0.5 & 58$^\circ\pm13^\circ$ & 333$\pm {61}$ & 0.8$\pm$0.3 & 0 & 2 (10.65) \\ % 1/1.1
\underline{2017-07-16} & 19:39:29 & $1.63\pm {0.19}$ & $1.40 \pm {0.12}$ & ${0.15} \pm {0.51}$ & 12$\pm {5}$ & 0.77 & 1.7$\pm$0.9 & 56$^\circ\pm15^\circ$ & 177$\pm {103}$ & 0.8$\pm$0.5 & 0 & 1 (11.26) \\ % 0.9/1.2
2017-10-21*$^{\dagger}$ & 03:24:57 & $2.16 \pm {0.34}$ & $2.48 \pm {0.57}$ & ${0.07} \pm {0.04}$ & 110$ \pm {8}$ & 1.05 & 5.4$\pm$1.0 & 77$^\circ\pm9^\circ$ & 493$\pm {37}$ & 2.0$\pm$0.4 & 2 (9.55) & 11 (213.64) \\ % 8.9/1.1
\underline{2018-01-17}* & 17:39:31 & $1.52 \pm {0.08}$ & $1.65 \pm {0.15}$ & ${0.20}\pm {0.34}$ & 34$ \pm {4}$ & 0.36 & 1.8$\pm$0.6 & 56$^\circ\pm9^\circ$ & 361 $\pm {39}$ & 0.7$\pm$0.3 & 7 (25.1) & 9 (44.26) \\ % 1.1/1.2 
\hline
\end{tabular}
\label{ShockParamAll}	%ShockParamRampTDS
\end{table}
\FloatBarrier

It is interesting to note that the event on 2017-05-22, which was the ramp capture with no waves $\geq 5$ mV/m detected in the frequency range of this study, had the highest $\beta$ of this subset (Table \ref{ShockParamAll}). The ramp TDS for this event, which triggered off a wave $\sim$0.5 s downstream of the ramp, covers only the latter half of the ramp. It is possible that the relatively high value of $\beta$ suppressed growth of high amplitude perturbations in the electric field, in addition to smaller free energy sources associated with the low Mach number. 

The TDS were more frequently transmitted from downstream of the shocks, suggesting the amplitude of waves tends to be higher in the downstream than in the upstream region (Table \ref{WavesAll}), since the transmission of TDS bursts is based upon amplitude trigger. Wave amplitudes are observed to reach higher amplitudes overall in supercritical shocks than in subcritical shocks, though this is not the case for every supercritical shock. Due to the function of the TDS capturing, the highest amplitude waves are saved with lower amplitude waves being discarded. From this, we can conclude that no waves (within the frequencies observed) of amplitude $\gtrsim 80$mV/m were observed in the subcritical shock ramps in this study.  While all wave modes being investigated in this study were observed downstream and in the ramp, there were few observations of the high frequency whistler mode waves upstream. This can be compared to \citet[]{Bren2010,Cattell2020}, who found whistler mode waves were associated with shocks, but did not distinguish between the upstream or downstream regions. In contrast, there is evidence for upstream low frequency whistler mode waves in 7 of the 12 events in our data set (Table \ref{LowUpWhistlerWaves}). Four of the five highest Mach number events ($M_f > 2$) did not have observed whistler precursors, nor did the event with the lowest Mach number. 
However, it has been shown previously that at supercritical shocks, these waves can be Doppler shifted out of the range of a fluxgate magnetometer \citep[]{W2012}. This is consistent with the results of \citet[]{W2017}, who found that $\sim$78\% of IP shocks had upstream whistlers with no dependence on shock parameters. Although earlier observations and simulations found ECDI waves only in the ramp or the foot region of the ramp, we observed only one ECDI wave in the ramp, and most were seen downstream. Determining the relative occurrence of different wave modes is difficult. However, when the number of packets was normalized to the time duration spent in each region, all wave modes were most commonly observed in the ramp, followed by downstream of the shocks, then upstream. The low number of ramp captures and lack of consistent, total ramp coverage limits the inferences that can be drawn from this, but the higher rate of incidence of downstream captures suggests large waves are more often present in downstream of the shocks than in the upstream region of the shock.

\begin{table}
\caption{A count of the number of large amplitude ($\geq$ 5 mV/m) wave packets of ion acoustic (IAW), ECDI, dispersive electrostatic (DEW), whistler mode (whistler), and electrostatic solitary waves. There were a total of 14 TDS with time upstream, 4 TDS with time in the ramp, and 56 TDS with time downstream over all events in this study. Note a single TDS could contain any, all, or none of these wave modes.}
\centering
\begin{tabular}{r c c c c c}% c
\hline
Location & IAW & ECDI & DEW & Whistler & Solitary \\%& TDS total \\
\hline
Upstream & 13 & 5 & 24 & 3 & 2 \\
Ramp & 6 & 1 & 3 & 1 & 3 \\
Downstream & 131 & 22 & 103 & 51 & 18 \\
\hline
\end{tabular}
\label{WavesAll}
\end{table}

%%%%%%%ONLY DATES W/UPSTREAM WHISTLERS%%%%%%%%%%%%%
\begin{table}
\caption{Upstream magnetic field low frequency whistler frequencies and amplitudes. Dates which had no clear evidence of upstream whistlers have been omitted.}
\centering
\begin{tabular}{c c c|c c c}
\hline
Date & Frequency (Hz) & Amplitude (nT) & Date & Frequency (Hz) & Amplitude (nT) \\
\hline
2011-07-23 & 0.2 & 1.3 & 2017-05-09 & 2.9 & 2.0 \\
2011-08-06 & 3.2 & 1.6 & 2017-06-20 & 2.9 & 0.4 \\
2011-09-19 & 1.7 & 0.4 & 2017-07-16 & 1.3 & 1.2 \\
2011-11-05 & 1.1 & 1.6 & 2018-01-17 & 0.9 & 1.2 \\
\hline
\end{tabular}
\label{LowUpWhistlerWaves}
\end{table}

In several events, such as 2011-09-08 and 2011-11-05, %(Figure \ref{DispersiveB}),
we observe electrostatic waves with frequency changes in time in both the increasing and decreasing senses, though not necessarily due to dispersion. One possibility is that frequency changes could be ion acoustic waves following the local ion plasma frequency through very small, $\lesssim$Debye-scale changes in density. The time resolution of the STEREO particle instruments does not allow us to determine whether there are changes in distributions on the short time scales of the wave frequency changes. Future studies, potentially using the S/WAVES Low Rate Science (LRS) data on STEREO (\cite{Boug2008}), or data from Parker Solar Probe, MMS, and ARTEMIS could investigate the association of the waves with density changes to test this idea. It is evident that in some cases the frequency changes may correspond to changes in the magnitude of the B field, however, this is not a consistent result. % (Figure \ref{DispersiveB}).
It was investigated whether the frequency change follows $nf_{ce}$ for integer $n$=1--16 harmonics, however the changes in $nf_{ce}$ are too small to account for the large frequency drift.

All wave modes were more often observed with at least one other mode present than by themselves. For waves that did appear alone, we analyzed the relationship between the amplitudes of these packets and $B_d/B_u$, $N_d/N_u$, change in thermal energy to change in kinetic energy ratio ($3k_b\Delta T / \Delta K$), $\beta$,  and $M_{f}$. There were only weak correlations for any wave mode analyzed with any of these parameters. Only looking at the largest wave amplitude in each event on each date, we do recover positive correlations between the largest waves and the Mach numbers in agreement with a previous study \citep[]{W2007}.

\section{Conclusions}
We have shown results of the first statistical study of quasi-perpendicular interplanetary shocks that shows the waves in the ramps vary dramatically both in amplitude and mode in the short (.1s) duration of the ramp. Within the ramps, as well as downstream of shocks, we see a significant number of large amplitude, ion acoustic-like, ECDI-like, electrostatic solitary, and high frequency whistler mode waves. In addition, we have presented the first observations of waves with properties similar to ion acoustic waves that have very rapid changes in frequency with time (``dispersive electrostatic waves"), and have shown that these waves are as prevalent as ion acoustic waves, which earlier studies at IP shocks have concluded were the dominant mode. Several case studies of crossings of the Earth's bow shock , utilizing waveform capture burst data covering the duration of the ramp, have described the variability of waves in the ramps of quasi-perpendicular shocks. These investigations, including \citet{H2006, Bren2013, Goodrich2018}, concluded that ion acoustic waves and/or ECDI waves were the dominant modes, but did not identify DEW.

The presence of electrostatic waves which exhibit a drift in their frequencies with time is very clear, though the mechanism controlling the frequency change has not been determined. Previous studies of waveform data do not have the burst duration needed to identify these waves. Similarly, filterbank data would average over these structures in frequency. Ongoing studies using MMS and ARTEMIS which have search coil data and provide higher resolution particle measurements could provide further insight to these wave modes and the possible mechanisms controlling their frequencies. Preliminary study of the ARTEMIS and MMS datasets for a few cases confirm the electrostatic nature of the waves. There also remain other wave modes which have yet to be identified, such as the left hand polarized wave observed during the 2017-10-21 event. Further studies utilizing the more complete data sets from MMS could provide more wave parameters and higher time resolution particle data could provide evidence of free energy sources available for generation of these waves. Data from interplanetary shocks at other distances from the sun (e.g., Parker Solar Probe or MAVEN) could also provide useful indicators about wave generation.

The results of this study support previous findings of the dominance of intermediate frequency waves (IAW, ECDI, and DEW) near interplanetary shocks, and in the ramp regions of shocks. This is the first study to track the evolution of waves throughout the shock ramp, and to be able to identify DEW, which are as common as IAW. The amplitude-based capture of waves suggests that waves in the regions just downstream of shocks ($<$ 1500 $\rho_{gi}$) usually have larger amplitudes than those upstream (with the exception of Langmuir waves), contributing to the energy carried in wave-particle interactions. We also observe ECDI waves in the downstream and upstream regions more often than within the foot and ramps, counter to the expectations from simulation.

Due to limitations in the STEREO particle instruments, detailed tests of proposed instability mechanisms for the waves observed in this study cannot be made. As shown above, wave modes often vary on time-scales of less than 0.1s and can as rapidly as 0.01s. Definitively determining the features that excite the different modes would require measurements of both ion and electron distributions on times scales on the order of 10s of milliseconds. In some cases, such as waves with harmonics at the electron cyclotron frequency and other features consistent with the electron cyclotron drift instability \citep{Bren2013}, we can infer that this mechanism, coupling of ion acoustic waves and electron Bernstein waves, is operating. However, theoretical studies suggested that the process would occur only in the foot and ramp regions \citep{ML2017}, due to the interaction of reflected ions and incident electrons. Although our study found waves in the ramp, the waves occurred primarily in the downstream region where different distributions than those seen in ramp and foot regions would be required to excite the waves.

The high-time resolution MMS measurements of particle distributions are more commensurate with the rapid variability of the wave modes. \citet{Goodrich2019} provided evidence that ion acoustic waves at a quasi-perpendicular bow shock were excited by bursts of reflected ions, with wave growth consistent with the impulsive nature of the observed waves. \citet{C2019} utilized MMS data at a quasi-perpendicular super-critical IP shock, and suggested that the small scale (solitary wave like) structures contributed to the cross-shock potential, and that large amplitude ion acoustic waves, observed in the ramp and downstream, were likely driven by either currents or the reflected ions, as suggested by \citet{Goodrich2018}. The observation of rapidly varying ion distributions at this IP shock suggests that, at least in some cases, the ion acoustic waves, the ECDI waves and possibly the ``dispersive'' waves are driven by these ions. \citet{Goodrich2019} also describe a wave-ion diffusion process that could contribute to the observed rapid wave variability, which is a potential mechanism for the variation we see at IP shocks.

The bursty nature of large amplitude waves, the variability of modes, and the frequent occurrence of multiple modes at interplanetary shocks indicates the complexity of shock physics. The large amplitudes, often with significant parallel electric fields, suggest that waves play an important role in shock dissipation and particle energization. Our results also provide strong evidence for the importance of obtaining high-time resolution particle measurements as well as long duration, high time resolution electric and magnetic field at interplanetary shocks.

\acknowledgments
Data is available from the STEREO website, and from CDAWEB. This paper uses data from the Heliospheric Shock Database, generated and maintained at the University of Helsinki. We thank Dr. Janet Luhmann and the IMPACT team at the University of California Berkeley, the PLASTIC Consortium, and the CDAWEB team at Goddard Space Flight Center for the use of their data and tools. The work was supported by the International Space Science Institute's (ISSI) International Teams programme. L.B.W. was partially supported by Wind MO\&DA grants and a Heliophysics Innovation Fund (HIF) grant. Research at UMN was supported by NASA grants NNX16AF80G, NNX14AK73G, and 80NSSC19K0305.

\bibliography{ApJ_Draft_Cohen_2020_Accepted_arxiv}

\begin{thebibliography}{}
\expandafter\ifx\csname natexlab\endcsname\relax\def\natexlab#1{#1}\fi
\providecommand{\url}[1]{\href{#1}{#1}}
\providecommand{\dodoi}[1]{doi:~\href{http://doi.org/#1}{\nolinkurl{#1}}}
\providecommand{\doeprint}[1]{\href{http://ascl.net/#1}{\nolinkurl{http://ascl.net/#1}}}
\providecommand{\doarXiv}[1]{\href{https://arxiv.org/abs/#1}{\nolinkurl{https://arxiv.org/abs/#1}}}

\bibitem[{Acu{\~n}a {et~al.}(2008)Acu{\~n}a, Curtis, Scheifele, Russell,
  Schroeder, Szabo, \& Luhmann}]{Acuna2008}
Acu{\~n}a, M.~H., Curtis, D., Scheifele, J.~L., {et~al.} 2008, Space Science
  Reviews, 136, 203, \dodoi{10.1007/s11214-007-9259-2}

\bibitem[{Andersson {et~al.}(2009)Andersson, Ergun, Tao, Roux, LeContel,
  Angelopoulos, Bonnell, McFadden, Larson, Eriksson, Johansson, Cully, Newman,
  Goldman, Glassmeier, \& Baumjohann}]{Andersson2009}
Andersson, L., Ergun, R.~E., Tao, J., {et~al.} 2009, Phys. Rev. Lett., 102,
  225004, \dodoi{10.1103/PhysRevLett.102.225004}

\bibitem[{Bale {et~al.}(1998)Bale, Kellogg, Larsen, Lin, Goetz, \&
  Lepping}]{Bale1998}
Bale, S.~D., Kellogg, P.~J., Larsen, D.~E., {et~al.} 1998, Geophysical Research
  Letters, 25, 2929, \dodoi{10.1029/98GL02111}

\bibitem[{Bale {et~al.}(2008)Bale, Ullrich, Goetz, Alster, Cecconi, Dekkali,
  Lingner, Macher, Manning, McCauley, Monson, Oswald, \& Pulupa}]{Bale2008}
Bale, S.~D., Ullrich, R., Goetz, K., {et~al.} 2008, The Electric Antennas for
  the STEREO/WAVES Experiment (New York, NY: Springer New York), 529--547.
\newblock \url{https://doi.org/10.1007/978-0-387-09649-0_17}

\bibitem[{Balikhin {et~al.}(2005)Balikhin, Walker, Treumann, Alleyne,
  Krasnoselskikh, Gedalin, Andre, Dunlop, \& Fazakerley}]{Balikhin2005}
Balikhin, M., Walker, S., Treumann, R., {et~al.} 2005, Geophysical Research
  Letters, 32, \dodoi{10.1029/2005GL024660}

\bibitem[{Bohm \& Gross(1949{\natexlab{a}})}]{B1949a}
Bohm, D., \& Gross, E.~P. 1949{\natexlab{a}}, Phys. Rev., 75, 1864,
  \dodoi{10.1103/PhysRev.75.1864}

\bibitem[{Bohm \& Gross(1949{\natexlab{b}})}]{B1949b}
---. 1949{\natexlab{b}}, Phys. Rev., 75, 1851, \dodoi{10.1103/PhysRev.75.1851}

\bibitem[{Bougeret {et~al.}(2008)Bougeret, Goetz, Kaiser, Bale, Kellogg,
  Maksimovic, Monge, Monson, Astier, Davy, Dekkali, Hinze, Manning,
  Aguilar-Rodriguez, Bonnin, Briand, Cairns, Cattell, Cecconi, Eastwood, Ergun,
  Fainberg, Hoang, Huttunen, Krucker, Lecacheux, MacDowall, Macher, Mangeney,
  Meetre, Moussas, Nguyen, Oswald, Pulupa, Reiner, Robinson, Rucker, Salem,
  Santolik, Silvis, Ullrich, Zarka, \& Zouganelis}]{Boug2008}
Bougeret, J.~L., Goetz, K., Kaiser, M.~L., {et~al.} 2008, Space Science
  Reviews, 136, 487, \dodoi{10.1007/s11214-007-9298-8}

\bibitem[{Breneman {et~al.}(2010)Breneman, Cattell, Schreiner, Kersten,
  Wilson~III, Kellogg, Goetz, \& Jian}]{Bren2010}
Breneman, A., Cattell, C., Schreiner, S., {et~al.} 2010, Journal of Geophysical
  Research: Space Physics, 115, \dodoi{10.1029/2009JA014920}

\bibitem[{Breneman {et~al.}(2013)Breneman, Cattell, Kersten, Paradise,
  Schreiner, Kellogg, Goetz, \& Wilson~III}]{Bren2013}
Breneman, A.~W., Cattell, C.~A., Kersten, K., {et~al.} 2013, Journal of
  Geophysical Research: Space Physics, 118, 7654, \dodoi{10.1002/2013JA019372}

\bibitem[{Cattell {et~al.}(2020)Cattell, Short, Breneman, \&
  Grul}]{Cattell2020}
Cattell, C.~A., Short, B., Breneman, A.~W., \& Grul, P. 2020, The Astrophysical
  Journal, 897, 126, \dodoi{10.3847/1538-4357/ab961f}

\bibitem[{Cohen {et~al.}(2019)Cohen, Schwartz, Goodrich, Ahmadi, Ergun,
  Fuselier, Desai, Christian, McComas, Zank, Shuster, Vines, Mauk, Decker,
  Anderson, Westlake, Le~Contel, Breuillard, Giles, Torbert, \& Burch}]{C2019}
Cohen, I.~J., Schwartz, S.~J., Goodrich, K.~A., {et~al.} 2019, Journal of
  Geophysical Research: Space Physics, 124, 3961, \dodoi{10.1029/2018JA026197}

\bibitem[{Coroniti(1970)}]{C1982}
Coroniti, F.~V. 1970, Journal of Plasma Physics, 4, 265,
  \dodoi{10.1017/S0022377800004992}

\bibitem[{Coroniti {et~al.}(1982)Coroniti, Kennel, Scarf, \& Smith}]{C1970}
Coroniti, F.~V., Kennel, C.~F., Scarf, F.~L., \& Smith, E.~J. 1982, Journal of
  Geophysical Research: Space Physics, 87, 6029,
  \dodoi{10.1029/JA087iA08p06029}

\bibitem[{Dum {et~al.}(1980)Dum, Marsch, \& Pilipp}]{Dum1980}
Dum, C.~T., Marsch, E., \& Pilipp, W. 1980, Journal of Plasma Physics, 23, 91,
  \dodoi{10.1017/S0022377800022170}

\bibitem[{Dyrud \& Oppenheim(2006)}]{DO2006}
Dyrud, L.~P., \& Oppenheim, M.~M. 2006, Journal of Geophysical Research: Space
  Physics, 111, \dodoi{10.1029/2004JA010482}

\bibitem[{Edmiston \& Kennel(1984)}]{EK1984}
Edmiston, J.~P., \& Kennel, C.~F. 1984, Journal of Plasma Physics, 32, 429,
  \dodoi{10.1017/S002237780000218X}

\bibitem[{Fairfield(1974)}]{F1974}
Fairfield, D.~H. 1974, Journal of Geophysical Research (1896-1977), 79, 1368,
  \dodoi{10.1029/JA079i010p01368}

\bibitem[{Fitzenreiter {et~al.}(2003)Fitzenreiter, Ogilvie, Bale, \&
  Vi{\~n}as}]{Fitz2003}
Fitzenreiter, R.~J., Ogilvie, K.~W., Bale, S.~D., \& Vi{\~n}as, A.~F. 2003,
  Journal of Geophysical Research: Space Physics, 108,
  \dodoi{10.1029/2003JA009865}

\bibitem[{Formisano \& Torbert(1982)}]{Formisano1982}
Formisano, V., \& Torbert, R. 1982, Geophysical Research Letters, 9, 207,
  \dodoi{10.1029/GL009i003p00207}

\bibitem[{Forslund {et~al.}(1972)Forslund, Morse, Nielson, \&
  Fu}]{Forslund1972}
Forslund, D., Morse, R., Nielson, C., \& Fu, J. 1972, The Physics of Fluids,
  15, 1303, \dodoi{10.1063/1.1694082}

\bibitem[{Fredricks {et~al.}(1970)Fredricks, Crook, Kennel, Green, Scarf,
  Coleman, \& Russell}]{F1968}
Fredricks, R.~W., Crook, G.~M., Kennel, C.~F., {et~al.} 1970, Journal of
  Geophysical Research (1896-1977), 75, 3751, \dodoi{10.1029/JA075i019p03751}

\bibitem[{Fredricks {et~al.}(1968)Fredricks, Kennel, Scarf, Crook, \&
  Green}]{F1970}
Fredricks, R.~W., Kennel, C.~F., Scarf, F.~L., Crook, G.~M., \& Green, I.~M.
  1968, Phys. Rev. Lett., 21, 1761, \dodoi{10.1103/PhysRevLett.21.1761}

\bibitem[{Fuselier \& Gurnett(1984)}]{FG1984}
Fuselier, S.~A., \& Gurnett, D.~A. 1984, Journal of Geophysical Research: Space
  Physics, 89, 91, \dodoi{10.1029/JA089iA01p00091}

\bibitem[{Galvin {et~al.}(2008)Galvin, Kistler, Popecki, Farrugia, Simunac,
  Ellis, M{\"o}bius, Lee, Boehm, Carroll, Crawshaw, Conti, Demaine, Ellis,
  Gaidos, Googins, Granoff, Gustafson, Heirtzler, King, Knauss, Levasseur,
  Longworth, Singer, Turco, Vachon, Vosbury, Widholm, Blush, Karrer, Bochsler,
  Daoudi, Etter, Fischer, Jost, Opitz, Sigrist, Wurz, Klecker, Ertl,
  Seidenschwang, Wimmer-Schweingruber, Koeten, Thompson, \&
  Steinfeld}]{Galvin2008}
Galvin, A.~B., Kistler, L.~M., Popecki, M.~A., {et~al.} 2008, Space Science
  Reviews, 136, 437, \dodoi{10.1007/s11214-007-9296-x}

\bibitem[{Gardner {et~al.}(1958)Gardner, Goertzel, Grad, Morawetz, Rose, \&
  Rubin}]{Gardner1958}
Gardner, C., Goertzel, H., Grad, H., {et~al.} 1958, Proceedings of the Second
  United Nations International Conference on the Peaceful Uses of Atomic
  Energy, 31, 230

\bibitem[{Gary {et~al.}(1975)Gary, Feldman, Forslund, \& Montgomery}]{Gary1975}
Gary, S.~P., Feldman, W.~C., Forslund, D.~W., \& Montgomery, M.~D. 1975,
  Journal of Geophysical Research (1896-1977), 80, 4197,
  \dodoi{10.1029/JA080i031p04197}

\bibitem[{Gary {et~al.}(1981)Gary, Gosling, \& Forslund}]{Gary1981}
Gary, S.~P., Gosling, J.~T., \& Forslund, D.~W. 1981, Journal of Geophysical
  Research: Space Physics, 86, 6691, \dodoi{10.1029/JA086iA08p06691}

\bibitem[{Giagkiozis {et~al.}(2018)Giagkiozis, Wilson, Burch, Le~Contel, Ergun,
  Gershman, Lindqvist, Mirioni, Moore, \& Strangeway}]{Giagkiozis2018}
Giagkiozis, S., Wilson, L.~B., Burch, J.~L., {et~al.} 2018, Journal of
  Geophysical Research: Space Physics, 123, 5435, \dodoi{10.1029/2018JA025343}

\bibitem[{Goodrich {et~al.}(2018)Goodrich, Ergun, Schwartz, Wilson~III, Newman,
  Wilder, Holmes, Johlander, Burch, Torbert, Khotyaintsev, Lindqvist,
  Strangeway, Russell, Gershman, Giles, \& Andersson}]{Goodrich2018}
Goodrich, K.~A., Ergun, R., Schwartz, S.~J., {et~al.} 2018, Journal of
  Geophysical Research: Space Physics, 123, 9430, \dodoi{10.1029/2018JA025830}

\bibitem[{Goodrich {et~al.}(2019)Goodrich, Ergun, Schwartz, Wilson~III,
  Johlander, Newman, Wilder, Holmes, Burch, Torbert, Khotyaintsev, Lindqvist,
  Strangeway, Gershman, \& Giles}]{Goodrich2019}
---. 2019, Journal of Geophysical Research: Space Physics, 124, 1855,
  \dodoi{10.1029/2018JA026436}

\bibitem[{Gurnett \& Anderson(1977)}]{GA1977}
Gurnett, D.~A., \& Anderson, R.~R. 1977, Journal of Geophysical Research
  (1896-1977), 82, 632, \dodoi{10.1029/JA082i004p00632}

\bibitem[{Gurnett {et~al.}(1979)Gurnett, Anderson, Tsurutani, Smith, Paschmann,
  Haerendel, Bame, \& Russell}]{G1979}
Gurnett, D.~A., Anderson, R.~R., Tsurutani, B.~T., {et~al.} 1979, Journal of
  Geophysical Research: Space Physics, 84, 7043,
  \dodoi{10.1029/JA084iA12p07043}

\bibitem[{Hess {et~al.}(1998)Hess, MacDowall, Goldstein, Neugebauer, \&
  Forsyth}]{H1998}
Hess, R.~A., MacDowall, R.~J., Goldstein, B., Neugebauer, M., \& Forsyth, R.~J.
  1998, Journal of Geophysical Research: Space Physics, 103, 6531,
  \dodoi{10.1029/97JA03395}

\bibitem[{Hoppe \& Russell(1983)}]{HR1983}
Hoppe, M.~M., \& Russell, C.~T. 1983, Journal of Geophysical Research: Space
  Physics, 88, 2021, \dodoi{10.1029/JA088iA03p02021}

\bibitem[{Hoppe {et~al.}(1982)Hoppe, Russell, Eastman, \& Frank}]{Hoppe1982}
Hoppe, M.~M., Russell, C.~T., Eastman, T.~E., \& Frank, L.~A. 1982, Journal of
  Geophysical Research: Space Physics, 87, 643, \dodoi{10.1029/JA087iA02p00643}

\bibitem[{Hull {et~al.}(2006)Hull, Larson, Wilber, Scudder, Mozer, Russell, \&
  Bale}]{H2006}
Hull, A.~J., Larson, D.~E., Wilber, M., {et~al.} 2006, Geophysical Research
  Letters, 33, \dodoi{10.1029/2005GL025564}

\bibitem[{Kanekal {et~al.}(2016)Kanekal, Baker, Fennell, Jones, Schiller,
  Richardson, Li, Turner, Califf, Claudepierre, Wilson~III, Jaynes, Blake,
  Reeves, Spence, Kletzing, \& Wygant}]{Kanekal2016}
Kanekal, S.~G., Baker, D.~N., Fennell, J.~F., {et~al.} 2016, Journal of
  Geophysical Research: Space Physics, 121, 7622, \dodoi{10.1002/2016JA022596}

\bibitem[{Kellogg(1962)}]{K1962}
Kellogg, P.~J. 1962, Journal of Geophysical Research (1896-1977), 67, 3805,
  \dodoi{10.1029/JZ067i010p03805}

\bibitem[{Kennel(1987)}]{K1987}
Kennel, C.~F. 1987, Journal of Geophysical Research: Space Physics, 92, 13427,
  \dodoi{10.1029/JA092iA12p13427}

\bibitem[{Kennel {et~al.}(1982)Kennel, Scarf, Coroniti, Smith, \&
  Gurnett}]{K1982}
Kennel, C.~F., Scarf, F.~L., Coroniti, F.~V., Smith, E.~J., \& Gurnett, D.~A.
  1982, Journal of Geophysical Research: Space Physics, 87, 17,
  \dodoi{10.1029/JA087iA01p00017}

\bibitem[{Koval \& Szabo(2008)}]{KS2008}
Koval, A., \& Szabo, A. 2008, Journal of Geophysical Research: Space Physics,
  113, \dodoi{10.1029/2008JA013337}

\bibitem[{Krasnoselskikh {et~al.}(2014)Krasnoselskikh, Balikhin, Walker,
  Schwartz, Sundkvist, Lobzin, Gedalin, Bale, Mozer, Soucek, Hobara, \&
  Comisel}]{Kr2013}
Krasnoselskikh, V., Balikhin, M., Walker, S.~N., {et~al.} 2014, The Dynamic
  Quasiperpendicular Shock: Cluster Discoveries (Boston, MA: Springer US),
  459--522.
\newblock \url{https://doi.org/10.1007/978-1-4899-7413-6_18}

\bibitem[{Krasnoselskikh {et~al.}(2002)Krasnoselskikh, Lemb{\`e}ge, Savoini, \&
  Lobzin}]{Kr2002}
Krasnoselskikh, V.~V., Lemb{\`e}ge, B., Savoini, P., \& Lobzin, V.~V. 2002,
  Physics of Plasmas, 9, 1192, \dodoi{10.1063/1.1457465}

\bibitem[{Lin {et~al.}(1998)Lin, Kellogg, MacDowall, Scime, Balogh, Forsyth,
  McComas, \& Phillips}]{L1998}
Lin, N., Kellogg, P.~J., MacDowall, R.~J., {et~al.} 1998, Journal of
  Geophysical Research: Space Physics, 103, 12023, \dodoi{10.1029/98JA00764}

\bibitem[{Lin {et~al.}(1995)Lin, Anderson, Ashford, Carlson, Curtis, Ergun,
  Larson, McFadden, McCarthy, Parks, R{\`e}me, Bosqued, Coutelier, Cotin,
  D'Uston, Wenzel, Sanderson, Henrion, Ronnet, \& Paschmann}]{L1995}
Lin, R.~P., Anderson, K.~A., Ashford, S., {et~al.} 1995, Space Science Reviews,
  71, 125, \dodoi{10.1007/BF00751328}

\bibitem[{Lin {et~al.}(2008)Lin, Curtis, Larson, Luhmann, McBride, Maier,
  Moreau, Tindall, Turin, \& Wang}]{L2008}
Lin, R.~P., Curtis, D.~W., Larson, D.~E., {et~al.} 2008, Space Science Reviews,
  136, 241, \dodoi{10.1007/s11214-008-9330-7}

\bibitem[{Luhmann {et~al.}(2008)Luhmann, Curtis, Schroeder, McCauley, Lin,
  Larson, Bale, Sauvaud, Aoustin, Mewaldt, Cummings, Stone, Davis, Cook,
  Kecman, Wiedenbeck, von Rosenvinge, Acuna, Reichenthal, Shuman, Wortman,
  Reames, Mueller-Mellin, Kunow, Mason, Walpole, Korth, Sanderson, Russell, \&
  Gosling}]{Luhmann2008}
Luhmann, J.~G., Curtis, D.~W., Schroeder, P., {et~al.} 2008, Space Science
  Reviews, 136, 117, \dodoi{10.1007/s11214-007-9170-x}

\bibitem[{Mangeney {et~al.}(1999)Mangeney, Salem, Lacombe, Bougeret, Perche,
  Manning, Kellogg, Goetz, Monson, \& Bosqued}]{M1999}
Mangeney, A., Salem, C., Lacombe, C., {et~al.} 1999, Annales Geophysicae, 17,
  307, \dodoi{10.1007/s00585-999-0307-y}

\bibitem[{Marcowith {et~al.}(2016)Marcowith, Bret, Bykov, Dieckman, Drury,
  Lemb{\`{e}}ge, Lemoine, Morlino, Murphy, Pelletier, Plotnikov, Reville,
  Riquelme, Sironi, \& Novo}]{M2016}
Marcowith, A., Bret, A., Bykov, A., {et~al.} 2016, Reports on Progress in
  Physics, 79, 046901, \dodoi{10.1088/0034-4885/79/4/046901}

\bibitem[{Matsukiyo \& Scholer(2006)}]{MS2006}
Matsukiyo, S., \& Scholer, M. 2006, Journal of Geophysical Research: Space
  Physics, 111, \dodoi{10.1029/2005JA011409}

\bibitem[{Morawetz(1961)}]{Mor1961}
Morawetz, C.~S. 1961, The Physics of Fluids, 4, 988, \dodoi{10.1063/1.1706449}

\bibitem[{{Mozer} {et~al.}(1978){Mozer}, {Torbert}, {Fahleson}, {Falthammar},
  {Gonfalone}, \& {Pedersen}}]{M1978}
{Mozer}, F.~S., {Torbert}, R.~B., {Fahleson}, U.~V., {et~al.} 1978, IEEE
  Transactions on Geoscience Electronics, 16, 258,
  \dodoi{10.1109/TGE.1978.294558}

\bibitem[{Muschietti \& Lemb{\`e}ge(2006)}]{ML2006}
Muschietti, L., \& Lemb{\`e}ge, B. 2006, Advances in Space Research, 37, 483 ,
  \dodoi{https://doi.org/10.1016/j.asr.2005.03.077}

\bibitem[{Muschietti \& Lemb{\`e}ge(2013)}]{ML2013}
---. 2013, Journal of Geophysical Research: Space Physics, 118, 2267,
  \dodoi{10.1002/jgra.50224}

\bibitem[{Muschietti \& Lemb\`ege(2017)}]{ML2017}
Muschietti, L., \& Lemb\`ege, B. 2017, Annales Geophysicae, 35, 1093,
  \dodoi{10.5194/angeo-35-1093-2017}

\bibitem[{Ogilvie {et~al.}(1977)Ogilvie, von Rosenvinge, \& Durney}]{O1977}
Ogilvie, K.~W., von Rosenvinge, T., \& Durney, A.~C. 1977, Science, 198, 131,
  \dodoi{10.1126/science.198.4313.131}

\bibitem[{Paschmann {et~al.}(1980)Paschmann, Sckopke, Asbridge, Bame, \&
  Gosling}]{P1980}
Paschmann, G., Sckopke, N., Asbridge, J., Bame, S., \& Gosling, J. 1980,
  Journal of Geophysical Research: Space Physics, 85, 4689,
  \dodoi{10.1029/JA085iA09p04689}

\bibitem[{Rodriguez \& Gurnett(1975)}]{RG1975}
Rodriguez, P., \& Gurnett, D.~A. 1975, Journal of Geophysical Research
  (1896-1977), 80, 19, \dodoi{10.1029/JA080i001p00019}

\bibitem[{Russell \& Hoppe(1983)}]{RH1983}
Russell, C.~T., \& Hoppe, M.~M. 1983, Upstream Waves and Particles (Dordrecht:
  Springer Netherlands), 155--172.
\newblock \url{https://doi.org/10.1007/978-94-009-7096-0_12}

\bibitem[{Sagdeev(1958)}]{S1958}
Sagdeev, R. 1958, in Plasma Physics and the Problem of Controlled Thermonuclear
  Research, Vol.~5, 454--460

\bibitem[{{Sagdeev}(1966)}]{S1966}
{Sagdeev}, R.~Z. 1966, Reviews of Plasma Physics, 4, 23

\bibitem[{Sauvaud {et~al.}(2008)Sauvaud, Larson, Aoustin, Curtis, M{\'e}dale,
  Fedorov, Rouzaud, Luhmann, Moreau, Schr{\"o}der, Louarn, Dandouras, \&
  Penou}]{S2008}
Sauvaud, J.~A., Larson, D., Aoustin, C., {et~al.} 2008, The IMPACT Solar Wind
  Electron Analyzer (SWEA) (New York, NY: Springer New York), 227--239.
\newblock \url{https://doi.org/10.1007/978-0-387-09649-0_9}

\bibitem[{Thomsen {et~al.}(1985)Thomsen, Gosling, Bame, \& Mellott}]{T1985}
Thomsen, M.~F., Gosling, J.~T., Bame, S.~J., \& Mellott, M.~M. 1985, Journal of
  Geophysical Research: Space Physics, 90, 137, \dodoi{10.1029/JA090iA01p00137}

\bibitem[{Tonks \& Langmuir(1929)}]{T1929}
Tonks, L., \& Langmuir, I. 1929, Phys. Rev., 33, 195,
  \dodoi{10.1103/PhysRev.33.195}

\bibitem[{Vasko {et~al.}(2018)Vasko, Mozer, Krasnoselskikh, Artemyev, Agapitov,
  Bale, Avanov, Ergun, Giles, Lindqvist, Russell, Strangeway, \&
  Torbert}]{Vasko2018}
Vasko, I.~Y., Mozer, F.~S., Krasnoselskikh, V.~V., {et~al.} 2018, Geophysical
  Research Letters, 45, 5809, \dodoi{10.1029/2018GL077835}

\bibitem[{Vedenov(1963)}]{V1963}
Vedenov, A.~A. 1963, Journal of Nuclear Energy. Part C, Plasma Physics,
  Accelerators, Thermonuclear Research, 5, 169,
  \dodoi{10.1088/0368-3281/5/3/305}

\bibitem[{Vi{\~n}as \& Scudder(1986)}]{VS1986}
Vi{\~n}as, A.~F., \& Scudder, J.~D. 1986, Journal of Geophysical Research:
  Space Physics, 91, 39, \dodoi{10.1029/JA091iA01p00039}

\bibitem[{{Wenzel} {et~al.}(1992){Wenzel}, {Marsden}, {Page}, \&
  {Smith}}]{W1992}
{Wenzel}, K.~P., {Marsden}, R.~G., {Page}, D.~E., \& {Smith}, E.~J. 1992,
  \aaps, 92, 207

\bibitem[{Wilson {et~al.}(2007)Wilson, Cattell, Kellogg, Goetz, Kersten,
  Hanson, MacGregor, \& Kasper}]{W2007}
Wilson, L.~B., Cattell, C., Kellogg, P.~J., {et~al.} 2007, Phys. Rev. Lett.,
  99, 041101, \dodoi{10.1103/PhysRevLett.99.041101}

\bibitem[{Wilson~III(2016)}]{W2016}
Wilson~III, L.~B. 2016, Low Frequency Waves at and Upstream of Collisionless
  Shocks (American Geophysical Union (AGU)), 269--291.
\newblock
  \url{https://agupubs.onlinelibrary.wiley.com/doi/abs/10.1002/9781119055006.ch16}

\bibitem[{Wilson~III {et~al.}(2009)Wilson~III, Cattell, Kellogg, Goetz,
  Kersten, Kasper, Szabo, \& Meziane}]{W2009}
Wilson~III, L.~B., Cattell, C.~A., Kellogg, P.~J., {et~al.} 2009, Journal of
  Geophysical Research: Space Physics, 114, \dodoi{10.1029/2009JA014376}

\bibitem[{Wilson~III {et~al.}(2010)Wilson~III, Cattell, Kellogg, Goetz,
  Kersten, Kasper, Szabo, \& Wilber}]{W2010}
---. 2010, Journal of Geophysical Research: Space Physics, 115,
  \dodoi{10.1029/2010JA015332}

\bibitem[{Wilson~III {et~al.}(2017)Wilson~III, Koval, Szabo, Stevens, Kasper,
  Cattell, \& Krasnoselskikh}]{W2017}
Wilson~III, L.~B., Koval, A., Szabo, A., {et~al.} 2017, Journal of Geophysical
  Research: Space Physics, 122, 9115, \dodoi{10.1002/2017JA024352}

\bibitem[{Wilson~III {et~al.}(2014{\natexlab{a}})Wilson~III, Sibeck, Breneman,
  Contel, Cully, Turner, Angelopoulos, \& Malaspina}]{W2014a}
Wilson~III, L.~B., Sibeck, D.~G., Breneman, A.~W., {et~al.} 2014{\natexlab{a}},
  Journal of Geophysical Research: Space Physics, 119, 6455,
  \dodoi{10.1002/2014JA019929}

\bibitem[{Wilson~III {et~al.}(2014{\natexlab{b}})Wilson~III, Sibeck, Breneman,
  Contel, Cully, Turner, Angelopoulos, \& Malaspina}]{W2014b}
---. 2014{\natexlab{b}}, Journal of Geophysical Research: Space Physics, 119,
  6475, \dodoi{10.1002/2014JA019930}

\bibitem[{Wilson~III {et~al.}(2012)Wilson~III, Koval, Szabo, Breneman, Cattell,
  Goetz, Kellogg, Kersten, Kasper, Maruca, \& Pulupa}]{W2012}
Wilson~III, L.~B., Koval, A., Szabo, A., {et~al.} 2012, Geophysical Research
  Letters, 39, \dodoi{10.1029/2012GL051581}

\bibitem[{Wilson~III {et~al.}(2013)Wilson~III, Koval, Szabo, Breneman, Cattell,
  Goetz, Kellogg, Kersten, Kasper, Maruca, \& Pulupa}]{W2013}
---. 2013, Journal of Geophysical Research: Space Physics, 118, 5,
  \dodoi{10.1029/2012JA018167}

\bibitem[{Zhang {et~al.}(1999)Zhang, Matsumoto, \& Kojima}]{Z1999}
Zhang, Y., Matsumoto, H., \& Kojima, H. 1999, Journal of Geophysical Research:
  Space Physics, 104, 28633, \dodoi{10.1029/1999JA900301}

\end{thebibliography}
\bibliographystyle{aasjournal}

\end{document}